\documentclass[aps,pre,twocolumn,showpacs,amsmath,amssymb]{revtex4-1} 
\usepackage{graphicx} 
\usepackage{dcolumn} 
\usepackage{bm} 

\begin{document}


\title{Approach and Coalescence of Liquid Drops in Air}

\author{Joseph D. Paulsen}
\email{paulsenj@uchicago.edu}
\affiliation{The James Franck Institute and Department of Physics, The University of Chicago, Chicago, IL 60637, USA}

\date{\today}

\begin{abstract}
The coalescence of liquid drops has conventionally been thought to have just two regimes when the drops are brought together slowly in vacuum or air: a viscous regime corresponding to the Stokes-flow limit and a later inertially-dominated regime. 
Recent work [Proc.\ Natl.\ Acad.\ Sci.\ {\bf 109}, 6857 (2012)] found that the Stokes-flow limit cannot be reached in the early moments of coalescence, because the inertia of the drops cannot be neglected then. 
Instead, the drops are described by an ``inertially limited viscous" regime, where surface tension, inertia, and viscous forces all balance. 
The dynamics continue in this regime until either viscosity or inertia dominate on their own. 
I use an ultrafast electrical method and high-speed imaging to provide a detailed description of coalescence near the moment of contact for drops that approach at low speed and coalesce as undeformed spheres. 
These measurements support a description of coalescence having three regimes. 
Signatures both before and after contact identify a threshold approach-speed for deformation of the drops by the ambient gas. 
\end{abstract}



\maketitle

\section{Introduction}

\begin{figure}[b] 
\centering 
\begin{center} 
\includegraphics[width=3.4in]{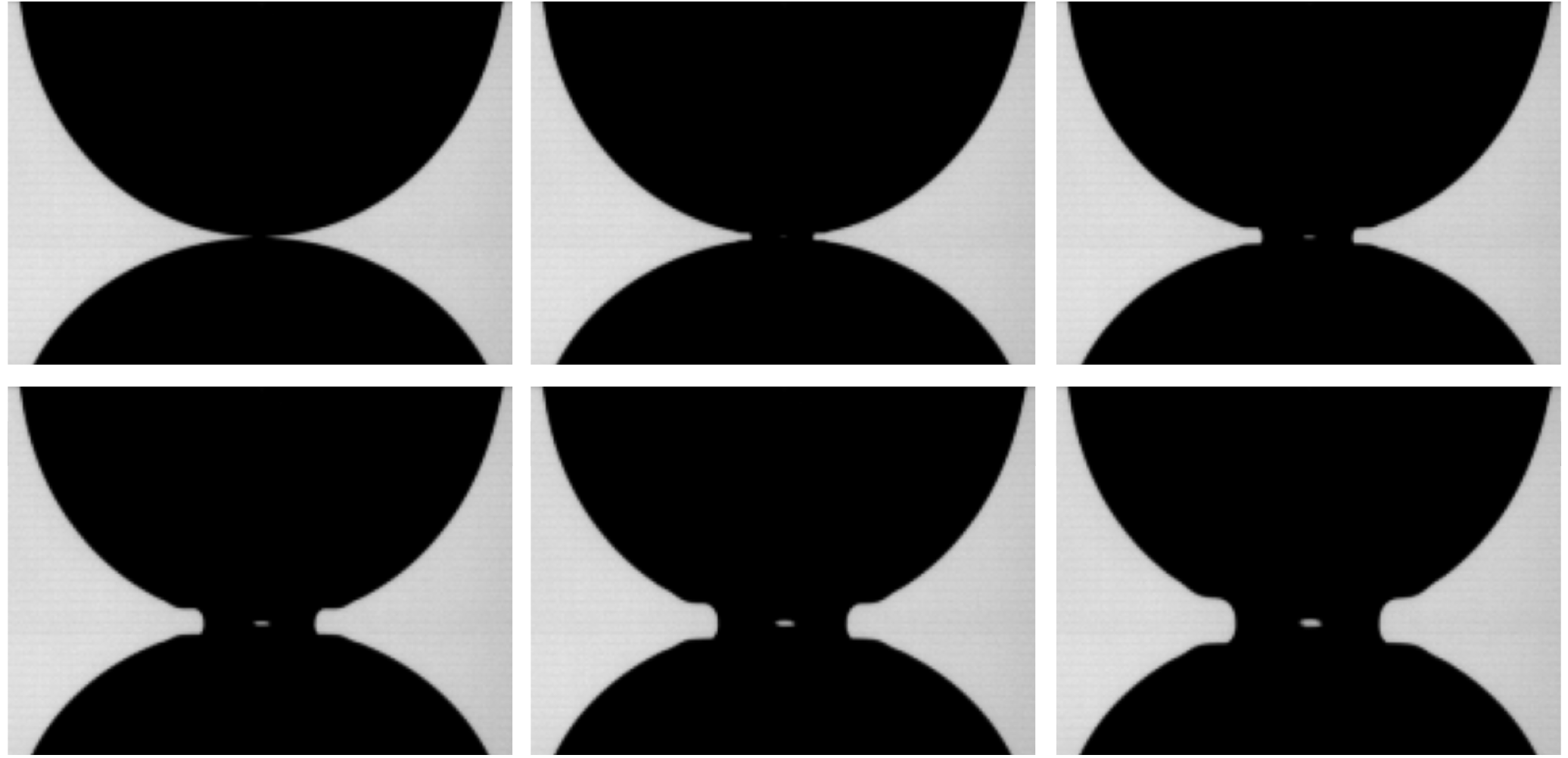} 
\end{center}
\caption{
Initial moments of coalescence for two water drops of radius $A=2$ mm in air. 
Frames are 120 $\mu$s apart. The central white spot is from the light source located behind the drops. 
A connecting neck of radius $r_{\text{min}}$ grows from infinitesimal size until it reaches the macroscopic size of the drops. 
}
\label{sequenceWater}
\end{figure}

When two liquid drops meet, a dramatic topological transformation takes place. 
A microscopic liquid neck forms between the drops, which rapidly expands due to surface tension forces (see Fig.\ \ref{sequenceWater}). 
This coalescence is ubiquitous in nature on a vast range of scales, from raindrops merging in clouds \cite{Pruppacher2010} to the coalescence of gas clouds during star formation \cite{Montmerle2006}. 
In industry, coalescence occurs in a multitude of situations, including viscous sintering \cite{Djohari2009}, oil desalting \cite{Eow2002}, dense spray systems \cite{Ashgriz1990}, and the processing and shelf-life of emulsions \cite{Evans1994}. 

The singularity inherent in drop coalescence has received significant attention. 
In the idealized limit where the drops form their initial contact over an infinitesimal region, the interfacial curvature, and therefore the pressure, diverge at the moment of contact, $t_0$. 
Although in reality this singularity is cut off (at least by the finite size of the molecules), it nonetheless controls the dynamics of the liquid neck over many orders of magnitude. 
These dynamics have recently been investigated by theoretical work \cite{Hopper1984,Hopper1990,Hopper1993a,Hopper1993b,Eggers1999,Eggers2003,Thompson2012_2}, numerical simulations \cite{Herrera1995,MenchacaRocha2001,Eggers2003,Lee2006,Thompson2012_2,Sprittles2012}, high-speed imaging experiments \cite{MenchacaRocha2001,Wu2004,Yao2005,Thoroddsen2005,Bonn2005,Burton2007,Yokota2011}, ultrafast x-ray phase-contrast imaging \cite{Fezzaa2008}, and an ultrafast electrical method \cite{Case2008,Case2009}. 
These studies sought to understand the dynamics governing the growth of the neck radius, $r_{\text{min}}$, as a function of time. 

\begin{figure}[bt] 
\centering 
\begin{center} 
\includegraphics[width=3.2in]{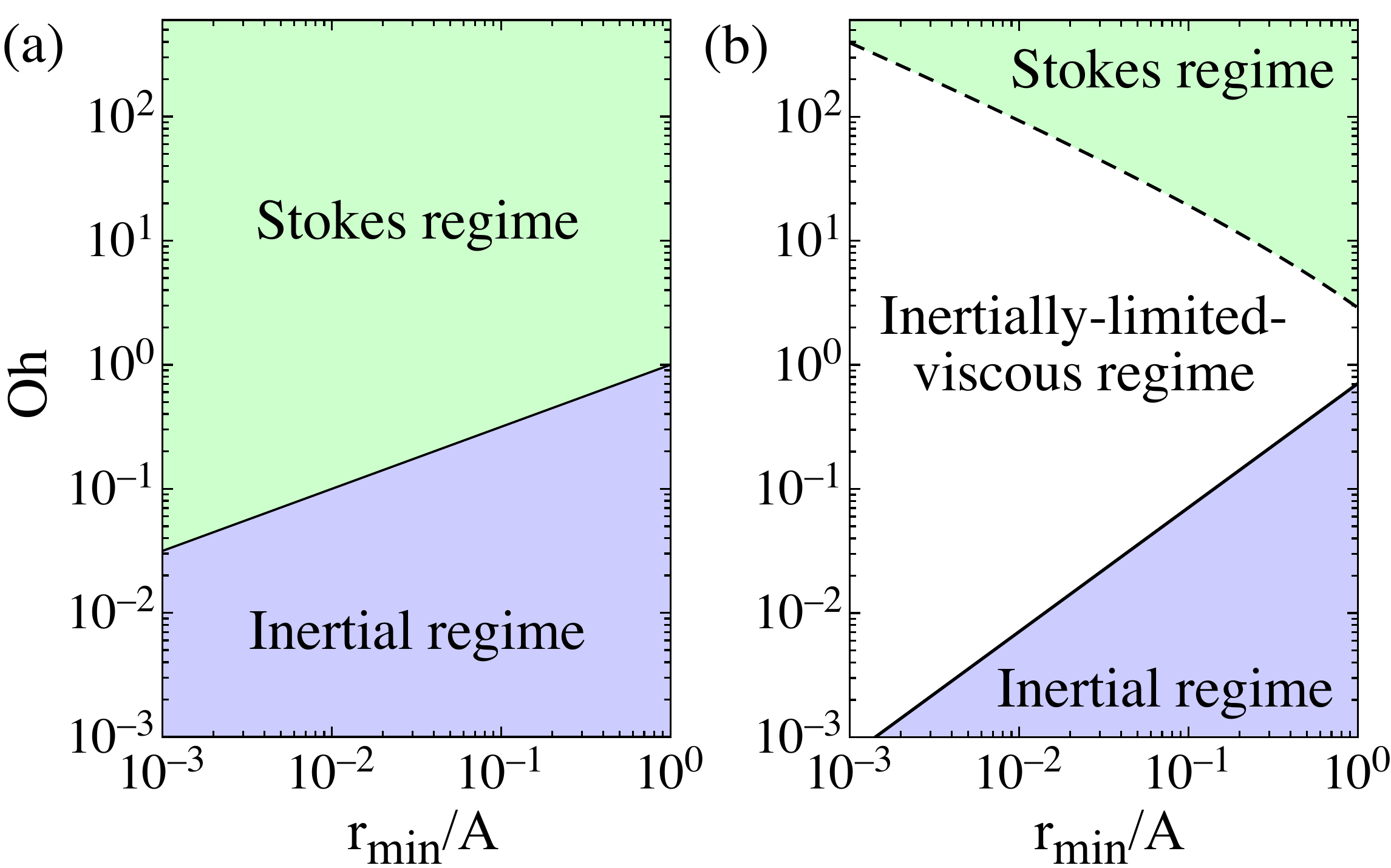}
\end{center}
\caption{
(Color online) 
Previous and revised phase diagram for drop coalescence in vacuum or air. 
Axes: non-dimensional neck radius, $r_{\text{min}}/A$, and $Oh=\mu/\sqrt{\rho \gamma A}$ (non-dimensional viscosity). 
(a) Old picture: there are two regimes, one dominated by viscous forces where inertia can be completely ignored (Stokes regime) and one dominated by inertia. 
Solid line: $r_{\text{min}}/A=Oh^2$. 
(b) Revised understanding: the ``inertially limited viscous" (ILV) regime intervenes at early times for finite-viscosity drops. 
The Stokes and inertial regimes do not share a phase boundary. 
Dashed line: Eq.\ (\ref{phase_boundary_3D}) with a proportionality constant of $1.4$. 
Solid line: Eq.\ (\ref{rcformula}). 
}
\label{phaseDiagram}
\end{figure}

The consensus from this work had been that there are just two regimes when the drops of radius $A$ are brought together in vacuum or air: (i) a highly viscous one dominated by macroscopic flows pulling the two drops together, and (ii) an inertial one described by local deformations near the growing neck. 
At early times, drops obey purely viscous behavior, and low-viscosity drops transition into the inertial regime later on. 
Thus, a coalescence phase diagram had been constructed \cite{Eggers1999}, which is shown in Fig.\ \ref{phaseDiagram}(a) in terms of the dimensionless neck radius, $r_{\text{min}}/A$, and the Ohnesorge number, $Oh=\mu/\sqrt{\rho \gamma A}$, which is a dimensionless viscosity (where $\mu$ is the dynamic viscosity, $\rho$ is the liquid density, and $\gamma$ is the interfacial tension). 

Recently, Paulsen \textit{et al.}\ \cite{Paulsen2012} demonstrated that a third regime---one that intervenes at early times for drops of any finite viscosity---had been missed. 
Using high-speed imaging experiments, an ultrafast electrical method, and high-accuracy computation, they showed that at small neck radius, the surface tension force pulling the drops together is too weak to overcome the inertia of the drops. 
Therefore, coalescence \textit{cannot} be in the Stokes-flow limit at early times, and an ``inertially limited viscous" regime occurs. 
In this regime, surface tension, viscous forces, and inertia all balance. 
At late times, viscous drops ($Oh>1$) \textit{can} reach the Stokes-flow limit, so there is an inertially limited viscous to Stokes crossover that is traversed as the neck radius grows. 

The final piece of the coalescence phase diagram is the viscous-to-inertial crossover time, $\tau_c$ (or the crossover radius, $r_c$), where the dynamics switch from the inertially limited viscous regime to a regime where only inertia is important. 
For many fluid flows, this crossover is easily identified by computing the dimensionless Reynolds number, $Re=\rho U L/\mu$, where $U$ and $L$ are characteristic velocity- and length-scales in the flows, respectively. 
Crossover behavior is expected when $Re\approx 1$. 
For coalescence, it was always assumed that $L=r_{\text{min}}$ \cite{Eggers1999,Eggers2003,Wu2004,Yao2005,Thoroddsen2005,Bonn2005,Burton2007,Case2008,Case2009,Thompson2012_2}. 

To observe the viscous-to-inertial crossover, Paulsen \textit{et al.}\ \cite{Paulsen2011} used an ultrafast electrical method (following refs.\ \cite{Case2008,Case2009}), which measures the neck radius down to tens of nanoseconds after the drops touch. 
For salt-water drops, Paulsen \textit{et al.}\ observed viscous behavior more than $3$ decades later than the prediction using the accepted Reynolds number for coalescence. 
To explain this discrepancy, they proposed that the dominant length-scale for the flows is instead given by the neck height, $L=r_{\text{min}}^2/A$. 
Thus, a revised phase diagram for coalescence was constructed, which is pictured in Fig.\ \ref{phaseDiagram}(b). 

This paper provides a more detailed experimental description and presents additional evidence for the picture developed in refs.\ \cite{Paulsen2012, Paulsen2011}. 
First, section \ref{ExperDesc} describes the electrical method, the fluids used, and the high-speed imaging technique, and section \ref{StokesInertialTheory} outlines several theoretical predictions for the purely viscous (Stokes) regime and the inertial regime. 

Section \ref{ILVregime} provides measurements and analysis of coalescence in the Stokes regime and the inertially limited viscous regime. 
Whereas Paulsen \textit{et al.}\ \cite{Paulsen2012} identified the inertially limited viscous to Stokes crossover by the motion of the back of the drops, I show that the same motions occur in the center-of-mass of the drops. 
The neck shapes in the inertially limited viscous and Stokes regimes are consistent with two distinct similarity solutions, and the interfacial curvature at the neck minimum can be used to distinguish between the regimes. 
The phase diagram is robust to different boundary conditions. 

Section \ref{VIcrossover} provides measurements and analysis of the viscous-to-inertial crossover. 
I collapse the electrical data with a different analysis from ref.\ \cite{Paulsen2011}, to demonstrate that the results are not sensitive to the details of the collapse protocol. 
I argue for a new Reynolds number for coalescence, as was done in ref.\ \cite{Paulsen2011}, now coming from the viscous side of the transition. 
I present high-speed imaging data where the surface tension is varied, which follows the crossover scaling calculated with the new Reynolds number. 

Section \ref{approach} studies the drops during their approach. 
Using optical, electrical resistance, and capacitance measurements, I show that at low approach-speed, the drops coalesce as undeformed spheres at finite separation. 
The data suggest that at low voltage, Van der Waals forces form the initial liquid neck (instead of forces due to the applied voltage).
The measurements provide an upper bound on the initial neck radius, $r_0$, which is smaller than previous estimates \cite{Thoroddsen2005,Fezzaa2008}. 


Appendix \ref{elecSystematic} reports checks on the electrical method, which show that the applied voltages and resulting electric fields do not affect the coalescence dynamics. Appendix \ref{prevCross} addresses previous measurements of the viscous-to-inertial crossover in the literature. 

This work gives a consistent picture wherein the inertially limited viscous regime is the asymptotic regime of liquid drop coalescence in vacuum or air. 
Viscous drops ($Oh>1$) transition into the Stokes regime later on, and low-viscosity drops ($Oh<1$) crossover into the inertial regime. 
In the inertially limited viscous regime and the inertial regime, the dominant flow gradients are on the scale of the neck height, $r_{\text{min}}^2/A$.

\section{Experimental Description}
\label{ExperDesc}

\subsection{Ultrafast electrical method}

In the experiment, two drops are formed on vertically aligned teflon nozzles of radius $A=2$ mm, which are separated by a distance $2A$. 
The pendant drop is fixed while the sessile drop is slowly grown with a syringe pump until the drops coalesce. 
Except for section \ref{approach}, the experiments are at sufficiently low approach-speed ($U_{\text{app}}< 9 \times 10^{-5}$ m/s) where the drops do not deform before contact.


\begin{figure}[bt] 
\centering 
\begin{center} 
\includegraphics[width=3.0in]{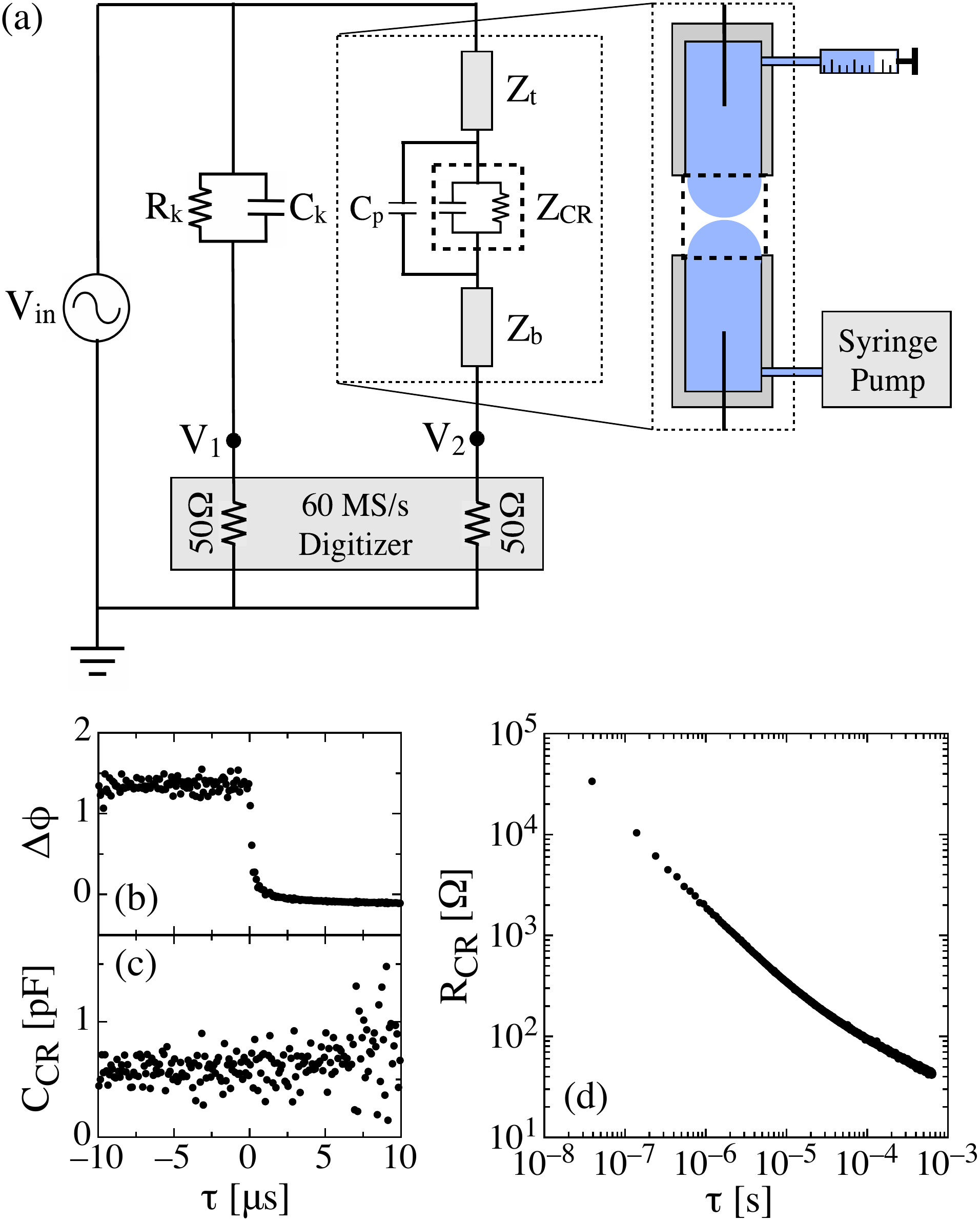}
\end{center}
\caption{
(Color online) 
Electrical method. 
(a) Coalescence cell and measurement circuit. 
Liquid hemispheres are formed on nozzles. 
One drop is grown slowly with a syringe pump (Razel Scientific, R-99) to initiate coalescence, while an AC voltage, $V_{\text{in}}$ (Hewlett-Packard, HP3325A), is applied across the drops and known circuit elements ($R_k$, $C_k$). 
Voltages $V_1$ and $V_2$ are recorded with a high-speed digitizer (NI PCI-5105, National Instruments) and converted to the time-varying complex impedance of the coalescence cell. 
$Z_{\text{CR}}$: impedance of the coalescence region (dashed box). 
$Z_t$, $Z_b$: impedances of the fluid-filled nozzles. 
$C_p$: stray capacitance between the nozzles. 
(b-d) Signals for a single saturated aqueous NaCl coalescence versus $\tau\equiv t-t_0$. 
(b) The phase angle, $\Delta\phi$, between $V_1$ and $V_2$ decreases sharply when the drops touch, which is used to measure $t_0$. 
(c) Capacitance of the coalescence region, $C_{\text{CR}}$, is roughly constant before and after contact. 
(d) Resistance of the coalescence region, $R_{\text{CR}}$, after contact.
}
\label{Schematic}
\end{figure}

Following the AC electrical method developed by Case \textit{et al.}\ \cite{Case2008,Case2009} and used in refs.\ \cite{Paulsen2012, Paulsen2011}, I measure the time-varying complex impedance, $Z_{\text{CR}}$, of two liquid hemispheres while they are coalescing (see Fig.\ \ref{Schematic}). 
Salt (NaCl) is added to the drops to make them electrically conductive. 
A high-frequency ($0.6 \leq f \leq 10$ MHz), low-amplitude ($V_{\text{in}} \leq 2$ V) AC voltage is applied across the drops by gold electrodes that are submerged in the fluid. 
By simultaneously sampling the voltage below the coalescence cell and the voltage below known passive circuit elements, the impedance of the coalescence cell is determined. 
Two backgrounds are subtracted: one is measured by bringing the nozzles together, and the second is a small parallel capacitance, $C_p=0.61\pm 0.12$ pF, that is measured before forming drops on the nozzles. 
This isolates the impedance of the coalescing drops, $Z_{\text{CR}}$, which is modeled as a time-varying resistor, $R_{\text{CR}}$, and capacitor, $C_{\text{CR}}$, in parallel. 
At the instant the drops touch, there is a sharp decrease in the phase difference, $\Delta\phi$, between the two measured voltages, which indicates the moment of contact, $t_0$, to within 1/$f$. 

Examples of these measured quantities are shown in Fig.\ \ref{Schematic}(b-d) as a function of $\tau \equiv t-t_0$, which measures time elapsed since the moment of contact, $t_0$. 
More than $10^4$ points are sampled, thereby capturing a large dynamic range from a single coalescence event. 

To ensure that the applied voltage and the resulting electric fields between the drops do not alter the coalescence dynamics, a variety of checks were performed on the electrical method (see Appendix \ref{elecSystematic}). 

\begin{figure}[bt] 
\centering 
\begin{center} 
\includegraphics[width=3.15in]{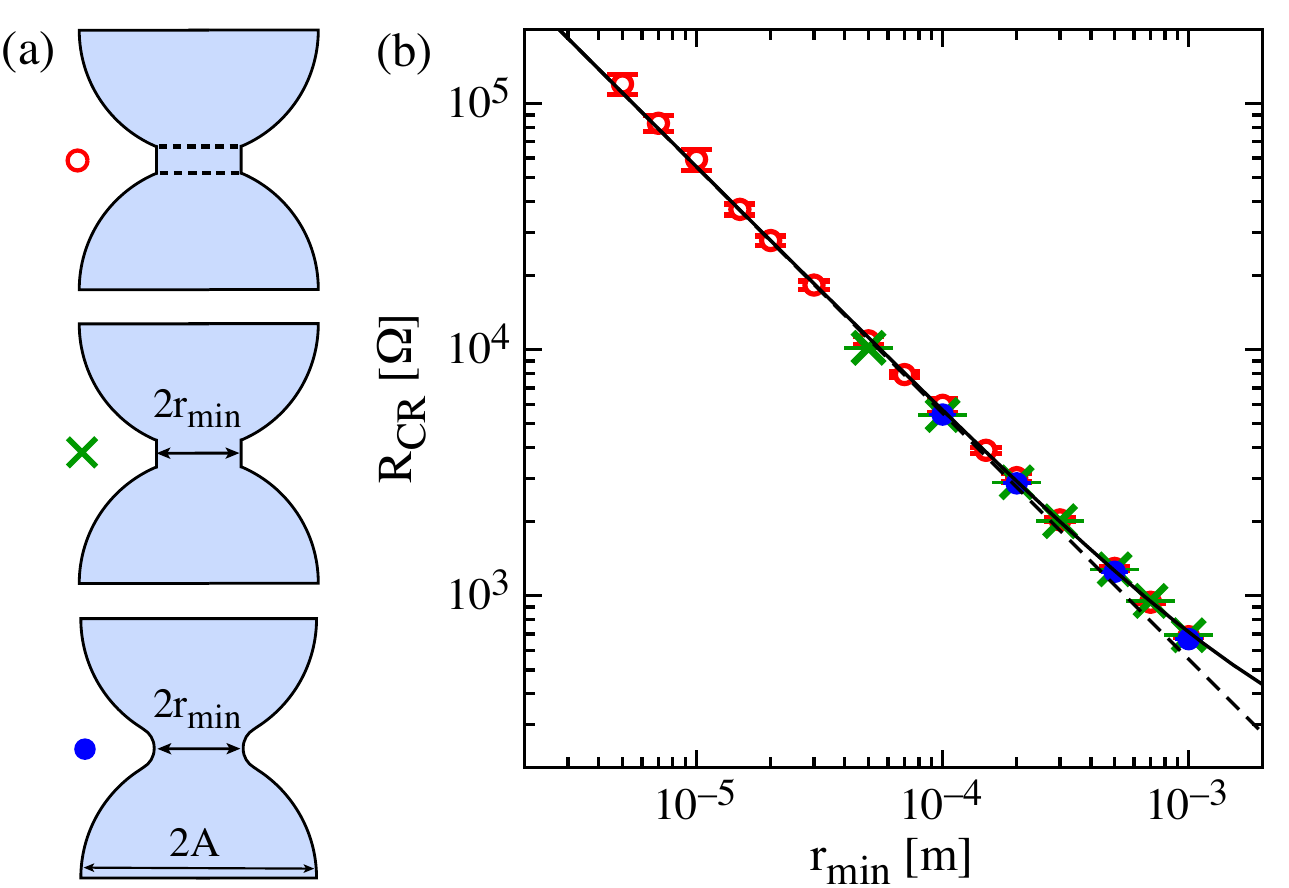} 
\end{center}
\caption{
(Color online) 
Conversion between electrical resistance, $R_{\text{CR}}$, and neck radius, $r_{\text{min}}$. 
(a) Three axisymmetric models of the coalescence region. 
Top to bottom: two truncated hemispheres are joined with a cylindrical neck of radius $r_{\text{min}}$, with planar equipotentials (dashed lines) sandwiching the neck; the same geometry without the equipotentials; two hemispheres joined smoothly with a circular arc. 
(b) Electrical resistance versus $r_{\text{min}}$ calculated numerically for $\sigma=1$ $\Omega^{-1}$m$^{-1}$ and $A=2$ mm. 
The data from all three models are well described by $R_{\text{CR}}=2/(\xi \sigma r_{\text{min}})+1/(\sigma \pi A)$ (solid line: Eq.\ (\ref{conversion})), where $\xi=3.62\pm 0.05$ is a fitting parameter. 
For small $r_{\text{min}}$, the data follow $R_{\text{CR}}=2/(\xi \sigma r_{\text{min}})$ (dashed line). 
}
\label{EStat}
\end{figure}

The conversion between $R_{\text{CR}}$ and $r_{\text{min}}$ is geometrical, and was determined numerically using the electrostatics calculation package EStat (FieldCo). 
To assess the dependance of the conversion on the choice of the model, this conversion was calculated in three different ways, pictured in Fig.\ \ref{EStat}(a). 
First, the conversion by Case \textit{et al.}\ \cite{Case2008,Case2009} was repeated, in which equipotentials are fixed on two planes that sandwich a cylindrical neck of radius $r_{\text{min}}$ and height $r_{\text{min}}^2/A$, so that the drops and their connecting neck are treated as series contributions to the total resistance. 
This calculation was compared with a second model with the same geometry but no such restriction on the field lines. 
In the third model, the shape of the interface is given by a circular arc connecting two hemispheres smoothly. 
As shown in Fig.\ \ref{EStat}(b), the three conversions agree within error bars, and the data are well described by:
\begin{equation}
R_{\text{CR}} = \frac{2}{\xi \sigma r_{\text{min}}} + \frac{1}{\sigma\pi A},
\label{conversion}
\end{equation}

\noindent where $\sigma$ is the electrical conductivity of the fluid and the dimensionless constant $\xi=3.62\pm 0.05$ is determined empirically. 

The first term in the conversion, $2/(\xi \sigma r_{\text{min}})$, is twice the resistance of a hemisphere with an opening of radius $r_{\text{min}}$. 
The constant term in the conversion, $1/(\sigma\pi A)$, can be understood as coming from the fluid neck itself. 
(This expression is the electrical resistance of a cylinder with radius $r_{\text{min}}$ and height $r_{\text{min}}^2/A$, with equipotentials on its flat faces.) 
Other neck geometries (e.g., an overturned neck shape, which is predicted to occur in the inertial regime \cite{Eggers2003, Fezzaa2008}) are expected to give the same conversion when $r_{\text{min}}$ is small, since the dominant term in the resistance comes from the general feature of a conducting hemisphere with an opening of radius $r_{\text{min}}$.

\subsection{Varying the liquid viscosity}

For the electrical measurements, the drops were mixtures of glycerol and water, with salt (NaCl) added to make the fluids electrically conductive. 
De-ionized water was saturated with NaCl at room temperature and mixed with glycerol. 
Each mixture was characterized by measuring its density, surface tension, viscosity, and electrical conductivity. 
Density was measured by weighing a known volume of fluid. 
Surface tension was measured by matching numerical solutions of the Young-Laplace equation to an image of a pendant drop. 
Viscosity was measured with glass capillary viscometers (Cannon-Fenske). 
Electrical conductivity was determined by measuring the electrical impedance of a thin cylindrical channel filled with fluid, using the coalescence cell and measurement circuit. 

\begin{figure}[bt] 
\centering 
\begin{center} 
\includegraphics[width=3.2in]{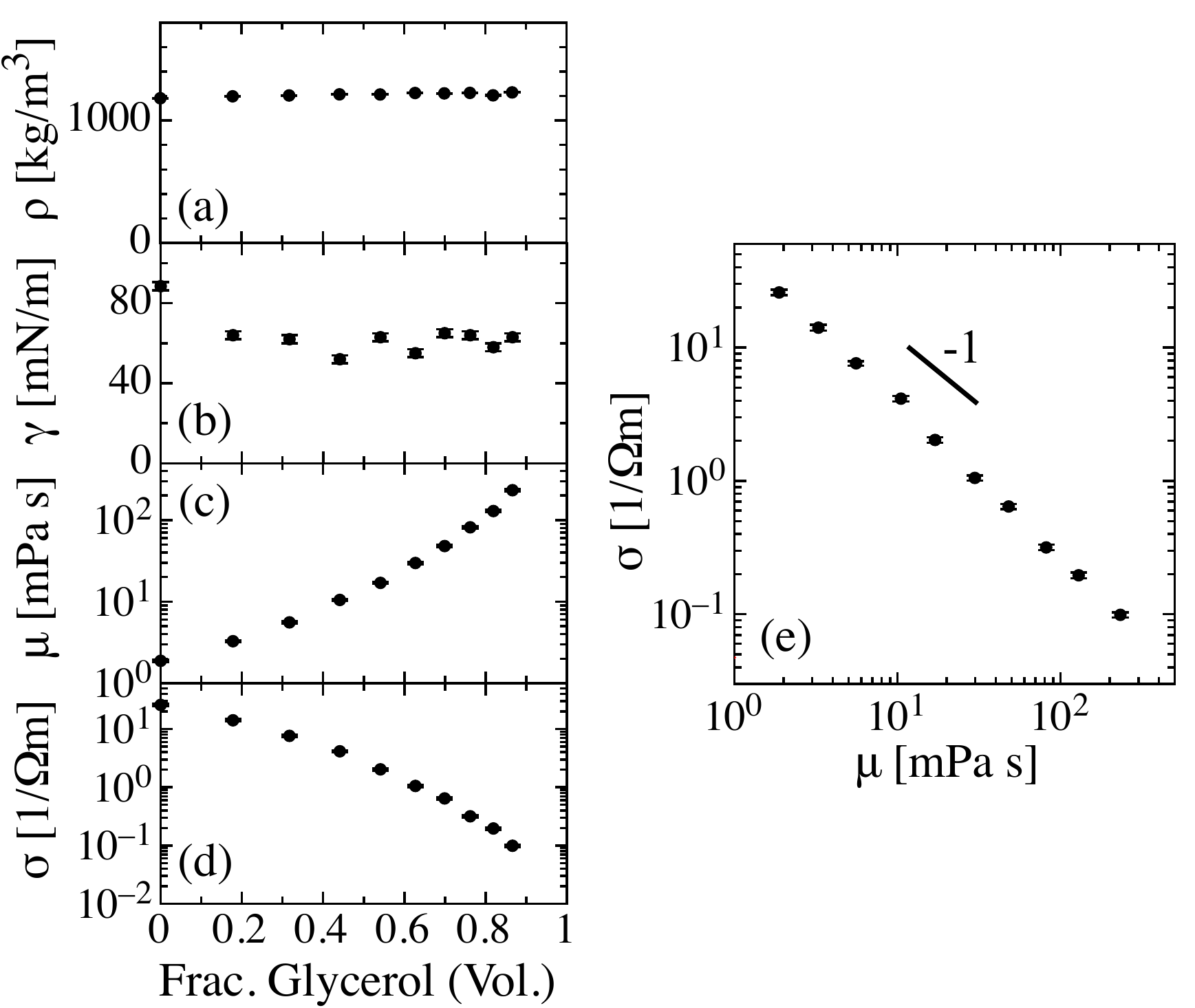} 
\end{center}
\caption{
Fluid parameters for glycerol-water-NaCl mixtures used for electrical measurements. 
(a) Mass density, $\rho$, is approximately constant over the range of mixtures used. 
(b) Surface tension, $\gamma$, is approximately constant. 
(Aqueous NaCl has $\gamma=88.5 \pm 2$ mN/m, which is higher than for pure water.) 
(c) Viscosity, $\mu$, varies over a large range. 
(d) AC electrical conductivity, $\sigma$ (at 1 to 10 MHz), decreases with increasing glycerol concentration. 
(e) AC electrical conductivity as a function of viscosity decreases slightly faster than $\mu^{-1}$. 
The low electrical conductivity at high viscosity sets the upper viscosity limit for the electrical method. 
}
\label{fluidParams}
\end{figure}

The measured fluid parameters are shown in Fig.\ \ref{fluidParams}. 
By changing the volume fraction of glycerol, the liquid viscosity was varied over two decades (from $1.9$ mPa s to $230$ mPa s) while the density and surface tension remained nearly constant, changing by factors of only $1.04$ and $1.6$, respectively. 

As shown in Fig.\ \ref{fluidParams}(e), the electrical conductivity decreases with increasing viscosity, which limits the experimental range of the viscosity of these mixtures with the electrical method. 
For a fixed, dilute concentration of NaCl, the relationship would obey: $\sigma\propto \mu^{-1}$. 
This expression comes from combining the Nernst-Einstein law (which relates conductivity to the ionic diffusion coefficients, $D$, at low ionic concentration: $\sigma\propto D$) with the Stokes-Einstein equation ($D\propto \mu^{-1}$). 
The conductivity falls off slightly faster than $\mu^{-1}$, which is consistent with the lower concentration of NaCl in the mixtures as the glycerol fraction is increased.
(There is another, smaller correction because the mixtures are not at low concentration, which has the opposite effect on the scaling.) 

\subsection{High-speed imaging}

A high-speed camera (Phantom v12, Vision Research) was used to observe other aspects of the coalescence dynamics, and to measure the neck radius versus time for silicone oils, which are non-conductive. 
The drops were precisely aligned with respect to the line-of-sight of the camera. 
Neck radii were measured using an edge-locating analysis on the images. 

\begin{figure}[bt] 
\centering 
\begin{center} 
\includegraphics[width=3.3in]{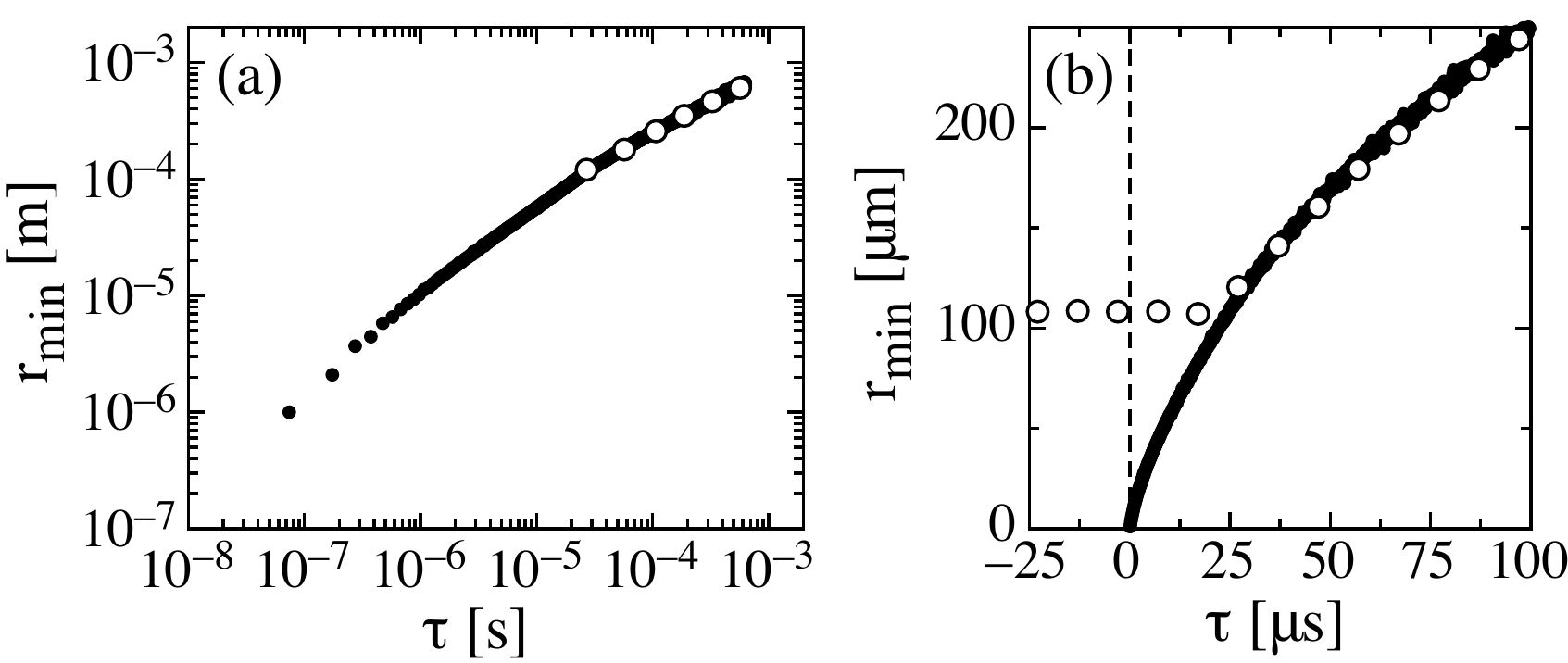} 
\end{center}
\caption{
Neck radius versus time for coalescing aqueous NaCl drops ($\mu=1.88$ mPa s, $\gamma=88.5$ mN/m, $\rho=1180$ kg/m$^3$). 
(a) Data from the electrical method ($\bullet$) and high-speed imaging ($\circ$), where $t_0$ is determined from a simultaneous electrical measurement. 
The two methods are in good agreement. 
The electrical data extends to far earlier times. 
(b) Data from the same experiments, on linear-linear axes, showing every camera frame. 
Before contact, the drop geometry and finite spatial resolution create an apparent neck of radius $110$ $\mu$m. 
The earliest imaging point that corresponds to the actual fluid neck is the third frame after $t_0$ ($\tau=27.0$ $\mu$s). 
}
\label{VideoElectric_rmin}
\end{figure}

To compare electrical measurements with high-speed imaging data, $r_{\text{min}}$ was measured both electrically and optically for saturated aqueous NaCl drops. 
For the optical data used in this comparison, $t_0$ was determined from a simultaneous electrical measurement, which was converted to the camera's time-base with a precision of $0.1$ $\mu$s. 
As shown in Fig.\ \ref{VideoElectric_rmin}(a), the two methods are in good agreement. 
The comparison serves as a quantitative check on the electrical method, and additionally illustrates the dynamic range gained by the electrical method versus high-speed imaging. 

In the current configuration, the dynamic range of high-speed imaging is determined by spatial resolution, as opposed to timing resolution. 
To see this, observe that imaging a neck of radius $r_{\text{min}}$ requires resolving a much smaller feature: the vertical gap between the drops, $r_{\text{min}}^2/A$. 
Thus, the minimum observable neck radius is set by the condition that $r_{\text{min}}^2/A$ is approximately equal to the spatial resolution of the optical setup (i.e., the neck height limits measurements of the neck width). 
For the experiment in Fig.\ \ref{VideoElectric_rmin}, the spatial resolution is 5.3 $\mu$m/pixel, so this estimate predicts that $r_{\text{min}}$ can be seen down to $100$ $\mu$m, which is consistent with the data. 
(To avoid this optical limitation, one can alternatively image \textit{through} the neck, as was done in recent drop spreading experiments \cite{Eddi2013_1}.) 

When comparing electrical and optical signals, a recent high-speed imaging study of coalescence reported a short delay ($20$ to $60$ $\mu$s) between the moment of electrical contact (from an electrical trigger for their ultrafast camera) and the first visible motion of the neck \cite{Thoroddsen2005}. 
The apparent delay between the electrical signal and visualized motion is now easily accounted for by the period of time when the neck height is smaller than the optical resolution. 
This explanation is also consistent with those authors' observation that the delay is shorter for smaller drops. 
To illustrate this point, Fig. \ref{VideoElectric_rmin}(b) compares electrical and optical measurements of the apparent neck size, $r_{\text{min}}$. 
Indeed, the early-time optical data give a constant value of $110$ $\mu$m, corresponding to the radius of the darkened region where the gap between the drops is smaller than the optical resolution. 

\section{Purely viscous (Stokes) and inertial regimes}
\label{StokesInertialTheory}

For purely viscous Stokes flow in two dimensions (2D), an exact analytic solution of coalescence was given by Hopper \cite{Hopper1984,Hopper1990,Hopper1993a,Hopper1993b}. 
The shape of the fluid interface at any instant during coalescence is an inverse ellipse, given parametrically by: 
\begin{subequations}
 \label{Hopper_contour}
 \begin{align}
  r(\theta) & = \sqrt{2}A\frac{(1-m^2)(1+m)\cos \theta}{\sqrt{1+m^2}(1+2m \cos 2\theta +m^2)}, \label{Hopper_contour_r} \\
  z(\theta) & = \sqrt{2}A\frac{(1-m^2)(1-m)\sin \theta}{\sqrt{1+m^2}(1+2m \cos 2\theta +m^2)}, \label{Hopper_contour_z}
 \end{align}
\end{subequations}

\noindent where $0\leq \theta <2\pi$, and the parameter $m$ is mapped to a neck radius by: 
\begin{equation}
r_{\text{min}}=A\sqrt{2}(1-m)/\sqrt{1+m^2}. 
\label{rmin_vs_m}
\end{equation}

\noindent This family of curves interpolates between two kissing circles ($m=1$) and a single circle ($m=0$). 
For small neck radius, these shapes limit to: 
\begin{equation}
(r^2+z^2)^2=4 A^2 z^2+r_{\text{min}}^2 r^2.
\label{inverse_ellipse}
\end{equation}

In the solution, the neck radius is given as a function of time by:
\begin{equation}
\frac{\gamma\tau}{\mu A} = \frac{\pi \sqrt{2}}{4} \int_{m^2}^1 \frac{ds}{s (1+s)^{1/2} K(s)},
\label{Hopper_exact}
\end{equation}

\noindent where $K(s)$ is the complete elliptic integral of the first kind, and $m$ is related to $r_{\text{min}}$ by Eq.\ (\ref{rmin_vs_m}). 
The asymptotic behavior of Eq.\ (\ref{Hopper_exact}) (in the limit that $r_{\text{min}}/A \rightarrow 0$) is given by the simple expression: 
\begin{equation}
\frac{\gamma\tau}{\mu A} = \frac{\pi r_{\text{min}}}{A} \left| \ln\left(\frac{r_{\text{min}}}{8 A}\right) \right|^{-1}.
\label{Hopper_approx}
\end{equation}

The early-time asymptotic form of 2D Stokes coalescence was extended to three dimensions (3D) by Eggers \textit{et al.}\ \cite{Eggers1999}. 
For asymptotically small neck radius,
\begin{equation}
r_{\text{min}} = \frac{\gamma\tau}{\pi\mu} \left|\ln\left({\frac{\gamma \tau}{\mu A}}\right)\right|,
\label{viscScaling}
\end{equation}

\noindent which they report is a reasonable approximation for $r_{\text{min}}\lesssim 0.03 A$. 



For inertially-dominated flows where the fluid viscosity is negligible, a scaling argument \cite{Eggers1999} predicted that in this regime,
\begin{equation}
r_{\text{min}} = D_0 \left( \frac{\gamma A}{\rho} \right)^{1/4} \tau^{1/2},
\label{invScaling}
\end{equation}

\noindent where $D_0$ is a dimensionless prefactor.
This scaling was seen in numerical simulations, which report $D_0=1.62$ \cite{Eggers2003}. 
High-speed imaging experiments \cite{MenchacaRocha2001, Wu2004, Thoroddsen2005, Bonn2005, Fezzaa2008} and other numerical simulations \cite{MenchacaRocha2001, Lee2006, Baroudi2014} have also observed this scaling regime and all report $D_0\approx 1$. 

\section{The inertially limited viscous (ILV) regime}
\label{ILVregime}



Recently, Paulsen \textit{et al.}\ \cite{Paulsen2012} showed that there is a third regime of liquid drop coalescence, which had been missed by previous experiments and was unanticipated by theory. 
The regime arises because the analytical Stokes solution cannot apply at early times, because it violates a simple force-balance when the neck is small. 
Namely, the macroscopic motion of the drops inherent in the Stokes solution requires a larger force than the vanishingly small neck can provide. 
Paulsen \textit{et al.}\ \cite{Paulsen2012} used simulation and experiment to show that at later times when the neck is larger, the Stokes regime is entered. 

Paulsen \textit{et al.}\ \cite{Paulsen2012} called this regime the ``inertially limited viscous" (ILV) regime, because the inertia of the drops prevents the Stokes solution from applying. 
In the ILV regime, the neck radius is empirically found to follow: 
\begin{equation}
r_{\text{min}} = C_0 \frac{\gamma}{\mu}\tau.
\label{ILV_Scaling}
\end{equation}
\noindent Previous experiments had observed this linear growth, but incorrectly assumed the drops to be in the Stokes regime \cite{Bonn2005,Thoroddsen2005,Burton2007,Paulsen2011,Yokota2011}. 

Paulsen \textit{et al.}\ \cite{Paulsen2012} used a force-balance argument to predict that for 3D drops, the Stokes regime is entered when: 
\begin{equation}
Oh \propto \left| \ln\left(\frac{1}{8}\frac{r_{\text{min}}}{A}\right)\right| \left(\frac{r_{\text{min}}}{A}\right)^{-1/2}.
\label{phase_boundary_3D}
\end{equation} 
Fig.\ \ref{phaseDiagram}(b) shows the phase diagram for liquid drop coalescence in 3D. 
The ILV regime occupies an increasingly larger portion of the phase-space as $r_{\text{min}}/A\rightarrow 0$. 
Thus, surface tension, inertia, and viscosity combine to form the true asymptotic regime of liquid drop coalescence. 
(However, if the drop viscosity is extremely large or small, the range where the ILV regime occurs may be below atomic scales, and so coalescence will start in the Stokes or the inertial regime.) 

In this section, I provide additional measurements and analysis of the ILV and Stokes regimes. 
These measurements support the new picture of coalescence developed by Paulsen \textit{et al.}\ \cite{Paulsen2012}.

\subsection{Change in velocity scaling}

Paulsen \textit{et al.}\ \cite{Paulsen2012} observed that the transition from the ILV regime to the Stokes regime would be accompanied by a change in the macroscopic velocity scaling of the drops. 
To observe this macroscopic motion, they used a geometry where two pendant drops hang from nozzles and are translated horizontally to initiate contact on their equators. 
Paulsen \textit{et al.}\ \cite{Paulsen2012} measured the velocity of a point on the back of one drop, $v_{\text{b.o.d.}}$, as a probe of the global motion of the drops, thus identifying the phase boundary between the ILV and Stokes regimes. 
Here, I measure the center-of-mass velocity of each drop, $v_{\text{c.o.m.}}$, and show that it gives consistent results. 

In the ILV regime, a force balance argument \cite{Paulsen2012} gives: 
\begin{equation}
v_{\text{c.o.m.}} \approx \frac{3\mu}{4A^3 \rho} r_{\text{min}}^2,
\label{vcom_ILV}
\end{equation}

\noindent In the Stokes regime, the 2D Stokes solution gives the asymptotic relationship: 
\begin{equation}
v_{\text{c.o.m.}} \approx \frac{\gamma}{2\pi\mu} \left(\frac{r_{\text{min}}}{A}\right) \left|\ln\left(\frac{1}{8}\frac{r_{\text{min}}}{A}\right)\right|,
\label{vcom_Stokes}
\end{equation}
\noindent which should apply for 3D drops as well \cite{Eggers1999}.

High-speed movies of the coalescing drops are analyzed to give the position of the center-of-mass of one drop, which is numerically differentiated to give $v_{\text{c.o.m.}}$, and averaged to suppress noise. 
(Because the movies only give the planar drop contour, I calculate the center-of-mass of the shape that is given by revolving the contour of the bottom half of one drop around the axis passing through the center of both drops.) 
The neck radius is measured directly from the same movie. 

\begin{figure}[tb]
\begin{center}
\includegraphics[width=2.8in]{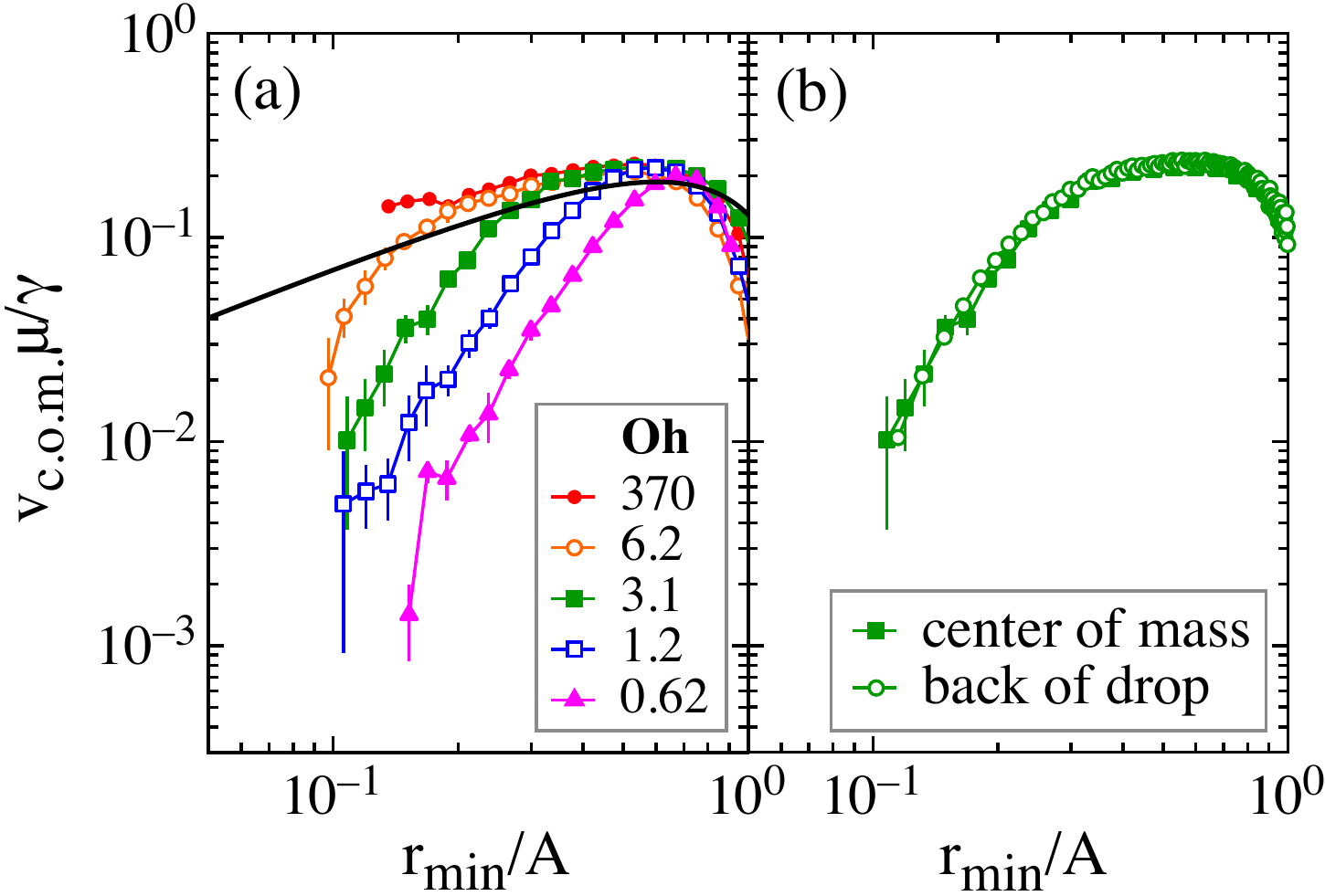}
\end{center}
\caption{
(Color online) 
ILV-to-Stokes crossover. 
(a) Rescaled center-of-mass velocity of the drops, $v_{\text{c.o.m.}}\mu/\gamma$, versus $r_{\text{min}}/A$ at several viscosities. 
Solid line: asymptotic result from 2D Stokes theory, Eq.\ (\ref{vcom_Stokes}). 
The data show super-linear growth at early times and merge onto the Stokes solution at late times. 
Higher-viscosity drops enter the Stokes regime at smaller neck radius. 
(b) The center-of-mass motion follows the motion of the back of one drop, here shown for $Oh=3.1$. 
}
\label{Vcom_vs_rmin}
\end{figure}

Figure \ref{Vcom_vs_rmin}(a) shows $v_{\text{c.o.m.}}$ rescaled by the viscous-capillary velocity, $\gamma/\mu$, versus the non-dimensional neck radius, $r_{\text{min}}/A$, for several viscosities. 
The data capture both an early dynamics where the global drop velocity is growing approximately as $r_{\text{min}}^2$ (as predicted for the ILV regime by Eq.\ (\ref{vcom_ILV})), and a late dynamics, where the data merge onto a master curve that is consistent with the Stokes theory, Eq.\ (\ref{vcom_Stokes}). 
The higher the fluid viscosity, the earlier the transition into the Stokes regime. 
In Fig.\ \ref{Vcom_vs_rmin}(b), $v_{\text{c.o.m.}}$ is shown for one of the viscosities along with $v_{\text{b.o.d.}}$ obtained from the same movie. 
The two measurements are in good agreement. 
This crossover in global drop motion marks the phase boundary between the ILV regime and the Stokes regime, which was reported in ref.\ \cite{Paulsen2012}, and is plotted in Fig.\ \ref{phaseDiagram}(b).

\subsection{Neck shapes}

The shape of the fluid neck connecting the coalescing drops offers another means of identifying the Stokes regime from the ILV regime. 
The neck shapes were compared by Paulsen \textit{et al.}\ \cite{Paulsen2012}, and a more detailed comparison is provided here. 

\begin{figure}[tb]
\begin{center}
\includegraphics[width=3.3in]{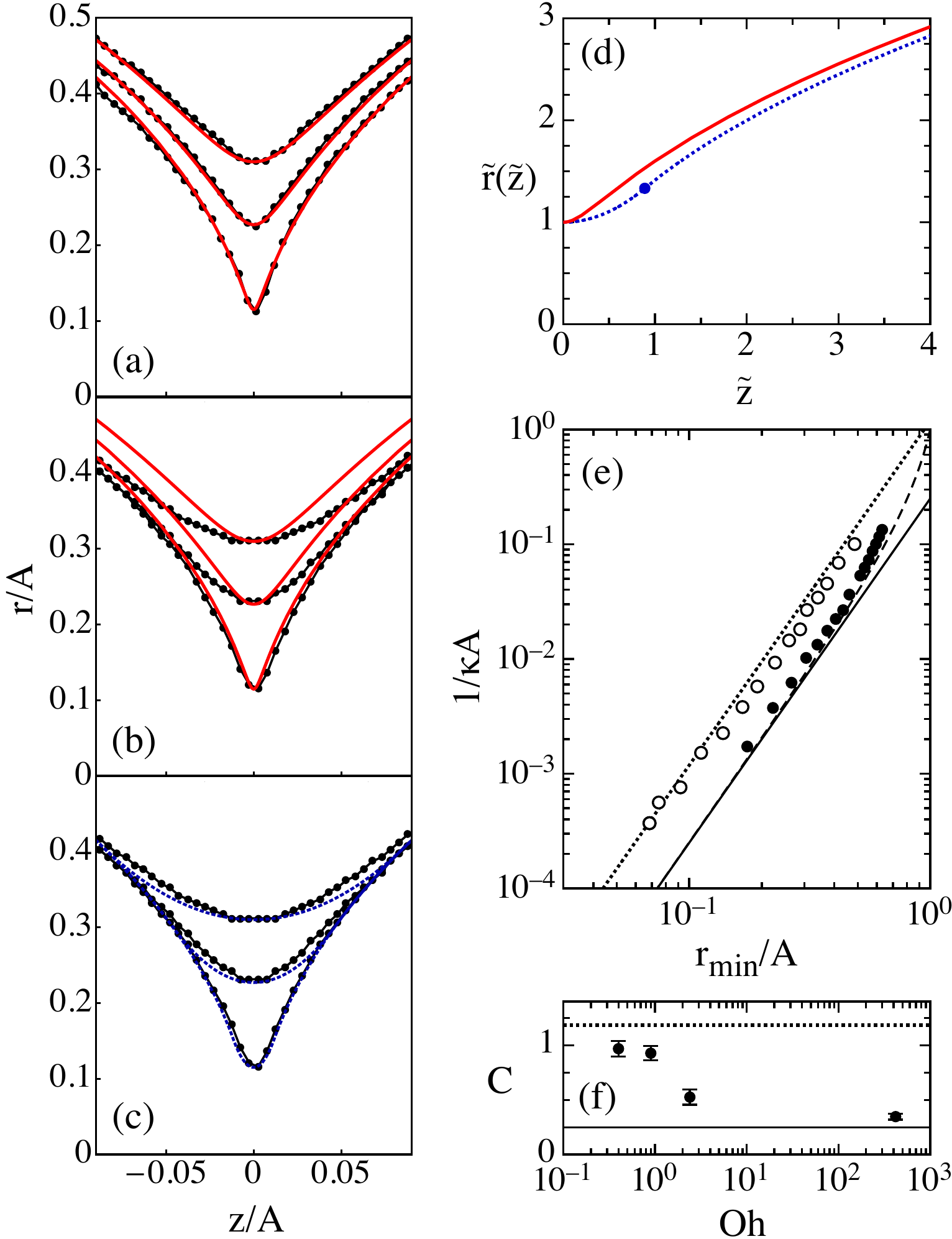}
\end{center}
\caption{
(Color online) 
Neck shape versus $r_{\text{min}}/A$ and viscosity. 
(a-c) Neck shape for high viscosity (a) ($\mu=58600$ mPa s, $Oh=370$) and intermediate viscosity (b,c) ($\mu=96.6$ mPa s, $Oh=0.62$). 
At high viscosity (a), the data agree with the exact Stokes-flow shapes (solid lines: Eq.\ (\ref{Hopper_contour})). 
At intermediate viscosity (b), the neck is much broader, and the Stokes theory is a poor fit to the data. 
Instead, the data is well described by two spheres joined smoothly with a parabolic neck (c). 
(d) Neck shape similarity solutions versus rescaled coordinates, $\tilde{r} \equiv r/r_{\text{min}}$, $\tilde{z} \equiv z/(r_{\text{min}}^2/A)$. 
Dotted line: parabolic neck connected to spherical drops, Eq.\ (\ref{simsol_Parabola}). 
Solid line: Stokes solution, Eq.\ (\ref{simsol_Hopper}). 
(e) Dimensionless radius of curvature at neck minimum, $1/\kappa A$, versus $r_{\text{min}}/A$. 
High-viscosity data ($\bullet$ $Oh=370$) agree with the Stokes theory (dashed line), which is approximated by Eq.\ (\ref{neck_curve}) with $C=1/4$ at early times (solid line). 
Intermediate-viscosity data ($\circ$ $Oh=0.62$) follow the result for a parabolic neck (dotted line: Eq.\ (\ref{neck_curve}) with $C=32/27$). 
(f) Curvature scaling prefactor, $C$, measured at fixed radius ($0.1<r_{\text{min}}/A<0.2$), versus $Oh$. 
As viscosity increases, the data transition from the value for a parabolic neck (dotted line: $32/27$) to the Stokes theory (solid line: $1/4$). 
}
\label{DropProfileFig}
\end{figure}

In Fig.\ \ref{DropProfileFig}(a) and (b), the exact shapes, Eq.\ (\ref{Hopper_contour}), are compared with experiments at $Oh=370$ and $Oh=0.62$. 
At each viscosity, the neck shapes are shown at three different times. 
To demonstrate that the regime can be identified by shape alone (without a knowledge of the time-dependance), I plot the shape that matches the minimum neck radius, $r_{\text{min}}$, of the experimental data. 
The Stokes solution agrees well with the high-viscosity data, where the drops have transitioned into the Stokes regime. 
However, it clearly fails to describe the shapes at $Oh=0.62$, where the drops are in the ILV regime, as shown in Fig.\ \ref{DropProfileFig}(b). 
In this regime, the neck is broader, perhaps because the drops have not translated towards each other appreciably. 
Instead, the neck shapes in this regime can be described by two kissing spheres joined smoothly by a parabolic neck, as shown in Fig.\ \ref{DropProfileFig}(c). 
(The parabola is uniquely determined by specifying $r_{\text{min}}$ and requiring that the parabolic region joins the two spheres continuously and with a continuous first derivative.)

On intermediate scales that are larger than the neck but smaller than the drops, the spherical drops are well-approximated by parabolas: $r=\sqrt{2Az}$ for $r_{\text{min}}\ll r \ll A$. 
A change of variables is made by rescaling the radial direction by the neck radius, $\tilde{r} \equiv r/r_{\text{min}}$, and the axial direction by the neck height, $\tilde{z} \equiv z/(r_{\text{min}}^2/A)$. 
This transformation collapses the profile onto itself: 
\begin{equation}
\tilde{r}(\tilde{z}) =
\begin{cases} 
1+\frac{27}{64}\tilde{z}^2, & \tilde{z}< \frac{8}{9}; \\
\sqrt{2 \tilde{z}}, & \tilde{z}\geq \frac{8}{9}. \\
\end{cases}
\label{simsol_Parabola}
\end{equation}

\noindent Thus, the parabolic-neck geometry is a similarity solution with radial scale $r_{\text{min}}$ and axial scale $r_{\text{min}}^2/A$. 
The solution is plotted in Fig.\ \ref{DropProfileFig}(d), with a point at $(\tilde{z},\tilde{r})=(8/9,4/3)$ marking where the neck merges onto the drops.
(The same rescaled coordinates, $r/r_{\text{min}}$ and $z/(r_{\text{min}}^2/A)$, were found to collapse the neck shapes in the inertial regime, for the case of neighboring spherical drops deposited on a substrate with a 90$^{\circ}$ contact angle \cite{Eddi2013_2}.) 

The drop shapes in the Stokes regime also form a similarity solution near the neck with the same scaling. 
As noted in ref.\ \cite{Eggers1999}, expansion of the drop contour near the neck for $z\ll r_{\text{min}}\ll A$ (using Eq.\ (\ref{Hopper_contour}) or Eq.\ (\ref{inverse_ellipse})) gives $r(z)=(\frac{1}{2}r_{\text{min}}^2+\sqrt{(\frac{1}{2}r_{\text{min}}^2)^2+4A^2z^2})^{1/2}$. 
Rescaling this shape yields: 
\begin{equation}
\tilde{r}(\tilde{z}) = \left(\frac{1}{2}+\sqrt{\frac{1}{4}+4\tilde{z}^2}\right)^{1/2},
\label{simsol_Hopper}
\end{equation}

\noindent which is plotted in Fig.\ \ref{DropProfileFig}(d). 

Although I have only identified the self-similarity of the drop \textit{shapes} during coalescence, it is possible that the flows share this self-similarity. 
If so, then radial flows will scale with the neck radius, $r_{\text{min}}$, and axial flows will scale with the neck height, $r_{\text{min}}^2/A$, in both the ILV and Stokes regimes. 
This further solidifies a claim argued in ref.\ \cite{Paulsen2011} and in section \ref{VIcrossover} that the flow scale in the axial direction is given by the neck height, $r_{\text{min}}^2/A$. 

The interfacial curvature in the $(z,r)$ plane at the neck minimum, $\kappa$, is a scalar quantity that distinguishes between the asymptotic neck shapes in the ILV and Stokes regimes. 
In both regimes, the similarity solutions obey: 
\begin{equation}
\frac{1}{\kappa A}=C \left(\frac{r_{\text{min}}}{A}\right)^3,
\label{neck_curve}
\end{equation}
 
\noindent where $C=\frac{32}{27}\approx 1.19$ in the ILV regime, and $C=\frac{1}{4}$ in the Stokes regime. 
These predictions are in good agreement with the data shown in Fig.\ \ref{DropProfileFig}(e) for the two regimes. 

Figure \ref{DropProfileFig}(f) shows $C$ at fixed radius ($0.1<r_{\text{min}}/A<0.2$) as a function of dimensionless viscosity, $Oh$. 
Near $Oh=1.5$, there is a transition between the values predicted for the Stokes and ILV regimes, marking the phase boundary at that radius.

\subsection{Effect of drop boundary condition}

In ref.\ \cite{Paulsen2012} and in this work, it has been shown that the Stokes and ILV regimes have different macroscopic drop motion. 
One natural question is whether the boundary condition on the drops can influence this motion and determine the coalescence regime (e.g., is the Stokes regime precluded if the drops are fixed to rigid objects?). 
In terms of the coalescence singularity, the crucial question is whether the boundary condition can affect the dynamics in the limit that $r_{\text{min}}/A\rightarrow 0$. 

\begin{figure}[tb]
\begin{center}
\includegraphics[width=3.2in]{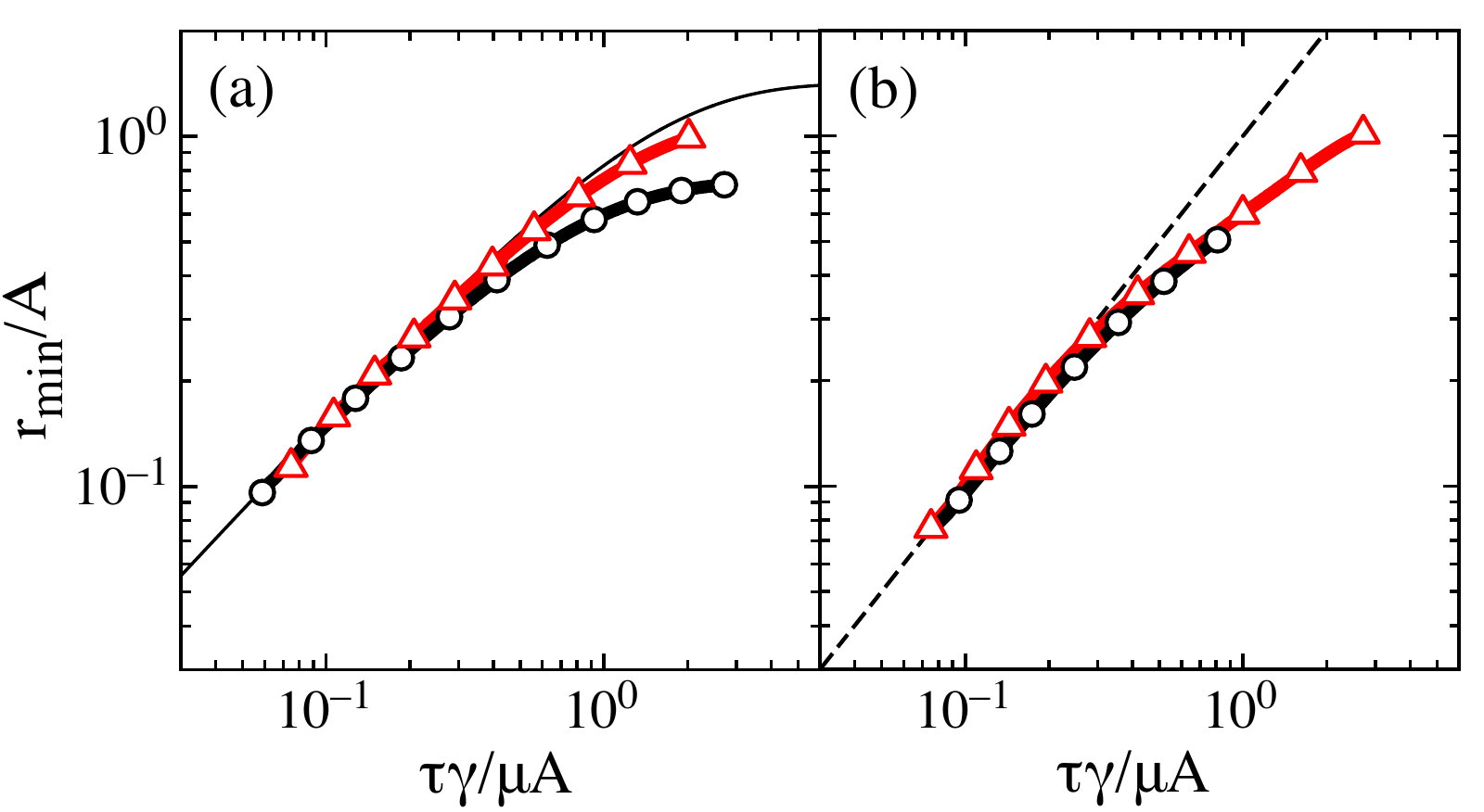}
\end{center}
\caption{
(Color online) Effect of drop boundary conditions.
(a) High-viscosity coalescence (silicone oil with $\mu=58600$ mPa s, $Oh=370$) for vertically-aligned hemispherical drops formed on nozzles ($\circ$) and for hanging pendant drops ($\triangle$). 
At early times, the neck growth does not depend on the boundary condition. 
The data follow the Stokes solution (solid line: Eq.\ (\ref{Hopper_exact})). 
(b) Intermediate-viscosity coalescence in these two drop geometries (glycerol-water-NaCl with $\mu=230$ mPa s ($\circ$) and $\mu=215$ mPa s ($\triangle$); $Oh\approx 1$).
The neck growth does not depend on the boundary condition. 
The data follow the ILV scaling (dashed line: Eq.\ (\ref{ILV_Scaling})). 
}
\label{NozzleGeomCompare}
\end{figure}

To address this issue, I compare measurements of $r_{\text{min}}$ under two contrasting boundary conditions: (i) the drops are hemispheres attached to fixed nozzles, and (ii) the drops are hanging and are coalesced at their equators, so that they can freely translate towards each other as they coalesce. 
I measure $r_{\text{min}}$ with both boundary conditions at high viscosity and intermediate viscosity. 

Figure \ref{NozzleGeomCompare}(a) compares high viscosity data with the 2D Stokes theory, Eq.\ (\ref{Hopper_exact}). 
(While the 2D and 3D theories match at early times, the 3D case does not make a prediction for $r_{\text{min}}> 0.03 A$, where all of the data lie. 
Therefore, following ref.\ \cite{Paulsen2012}, the data are compared with the 2D exact analytic solution.) 
The data follow the theory for small $r_{\text{min}}/A$; only at later times do the curves begin to depart from each other. 
Figure \ref{NozzleGeomCompare}(b) shows that for intermediate-viscosity drops, the boundary condition has a negligible effect on the dynamics, and the data matches the neck scaling for the ILV regime, Eq.\ (\ref{ILV_Scaling}). 
Thus, the phase diagram shown in Fig.\ \ref{phaseDiagram}(b) applies to coalescing drops that are fixed or free.

\section{Viscous-to-inertial crossover}
\label{VIcrossover}

Thus far, I have reported coalescence measurements in the Stokes and the ILV regimes, and I have observed the ILV-to-Stokes crossover by measuring the macroscopic motion of the drops. 
The remaining component of the coalescence phase diagram is the viscous-to-inertial crossover (from the ILV regime to the inertial regime). 

Recently, Paulsen \textit{et al.}\ \cite{Paulsen2011} used an ultrafast electrical method to measure this crossover for salt-water drops, and reported a major discrepancy with the theory \cite{Eggers1999,Eggers2003}. 
Whereas the theory predicts a crossover time between these regimes of $t_c \approx 0.7$ ns, the experiments show $t_c \approx 2$ $\mu$s. 
In terms of the neck size, the crossover radius was predicted to be $r_c \approx 30$ nm, whereas experiment showed $r_c \approx 20$ $\mu$m. 

To investigate this discrepancy, experiments were carried out where the liquid viscosity was varied over a large range \cite{Paulsen2011}. 
The data was found to be consistent with a newly proposed Reynolds number for coalescence, which is based on a smaller length scale for the dominant flow gradients given by the neck height, $L=r_{\text{min}}^2/A$. 
Here, I provide additional measurements and analysis of the viscous-to-inertial crossover, which support the conclusions of ref.\ \cite{Paulsen2011}.

\subsection{Collapse of electrical data}
\label{PrelimComp}

\begin{figure}[bt]
\centering 
\begin{center} 
\includegraphics[width=3.2in]{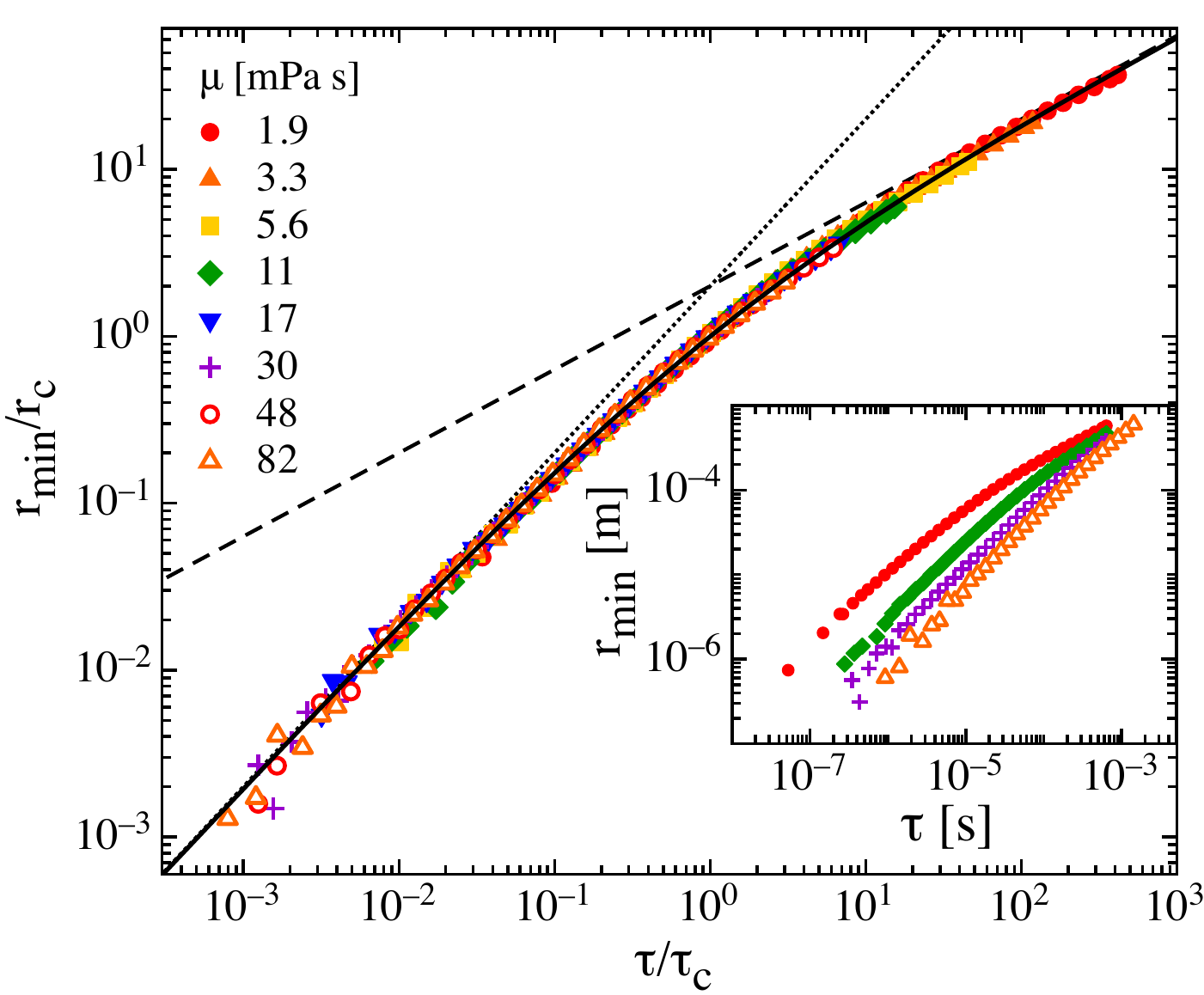}
\end{center}
\caption{
(Color online) 
\textit{Inset}: Neck radius versus time for glycerol-water-NaCl mixtures of different viscosities, from $1.9$ to $82$ mPa s. 
At each viscosity, data from 5 or more coalescence events are logarithmically binned and averaged. 
\textit{Main}: The data is collapsed by rescaling the x- and y-axes. 
Rescaling parameters $\tau_c$ and $r_c$ are obtained for each viscosity by fitting the data to Eq.\ (\ref{interpolate}) (solid line). 
The collapsed data and the fit exhibit asymptotic behavior of $2\tau/\tau_c$ (dotted line) at early times and $2\sqrt{\tau/\tau_c}$ (dashed line) at late times. 
} 
\label{muCollapse}
\end{figure}

The inset to Fig.\ \ref{muCollapse} shows $r_{\text{min}}$ versus time for 4 viscosities, ranging from $1.9$ to $82$ mPa s, which were measured electrically. 
In ref.\ \cite{Paulsen2011}, these data were rescaled to fall onto a master plot by rescaling the vertical and horizontal axes with free parameters, $r_c$ and $\tau_c$, at each viscosity to produce the best collapse. 
I perform a different analysis here, in order to demonstrate that the results are not sensitive to the particular way in which the data collapse is obtained. 

In ref.\ \cite{Paulsen2011}, after collapsing the data, it was noted that all of the data followed the simple interpolation:
\begin{equation}
\frac{r_{\text{min}}}{r_c}=2 \left(\frac{1}{\tau/\tau_c} + \frac{1}{\sqrt{\tau/\tau_c}}\right)^{-1}.
\label{interpolate}
\end{equation}

\noindent Here, I start with Eq.\ (\ref{interpolate}) and use it to collapse the data. 
For each viscosity, Eq.\ (\ref{interpolate}) is fit to the data, where $r_c$ and $\tau_c$ are fitting parameters. 
The data is then rescaled by the $r_c$ and $\tau_c$. 

\begin{figure}[bt] 
\centering 
\begin{center}
\includegraphics[width=3.4in]{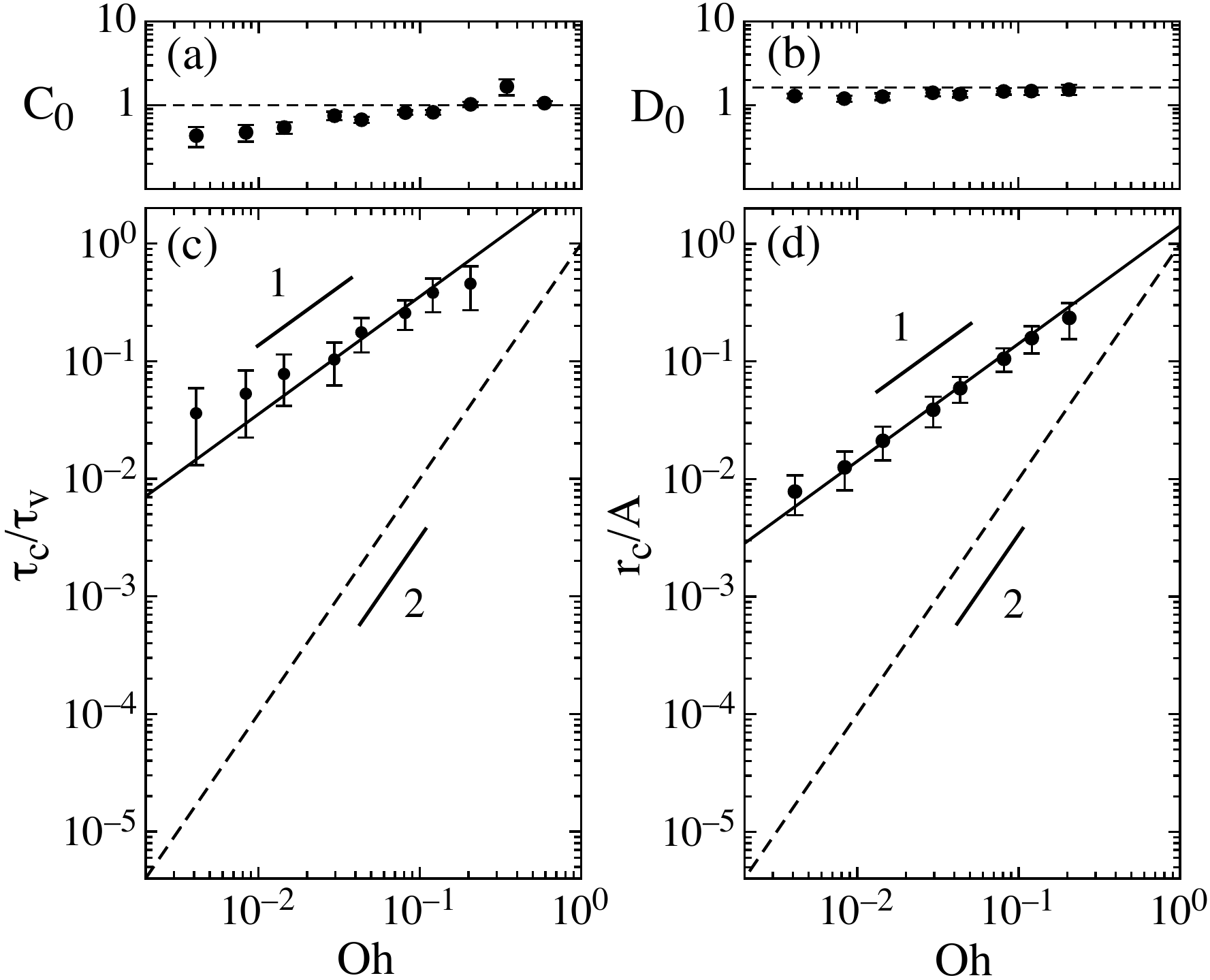} 
\end{center}
\caption{
(a,b) Measured dimensionless scaling-law prefactors, $C_0$ and $D_0$, versus $Oh$. 
In (a), the dashed line is $C_0=1$. 
In (b), the dashed line is the value from simulation \cite{Eggers2003}: $D_0=1.62$. 
(c) Rescaled viscous-to-inertial crossover time versus $Oh$. 
The dashed line shows $\tau_c/\tau_{\text{v}} = Oh^2$ (where $\tau_{\text{v}}= \mu A/\gamma$ is the viscous timescale), as predicted in the literature \cite{Eggers1999, Eggers2003}. 
Clearly this is a poor description of the data. 
The crossover radius proposed by ref.\ \cite{Paulsen2011} (with $\tau_c/\tau_{\text{v}} \propto Oh$) is consistent with the data (solid line: Eq.\ (\ref{ourtc})). 
(d) Rescaled viscous-to-inertial crossover radius versus $Oh$. 
The dashed line shows $r_c/A = Oh^2$, which was proposed in the literature \cite{Eggers1999, Eggers2003}. 
This fails to capture the data. 
The crossover radius proposed in this work describes the data well, with $r_c/A \propto Oh$ (solid line: Eq.\ (\ref{rcformula})). 
In (a-d), the error bars are determined by the fits to Eq.\ (\ref{interpolate}). 
} 
\label{rcvsOh}
\end{figure}

Figure \ref{muCollapse} shows the collapsed data, which fall cleanly onto a single curve given by Eq.\ (\ref{interpolate}). 
The scaling parameters, $r_c$ and $\tau_c$, determine the coefficients for the early- and late-time scaling laws, $C_0$ and $D_0$ (defined by eqns.\ \ref{ILV_Scaling} and \ref{invScaling}, respectively). 
Figure \ref{rcvsOh}(a) and (b) shows these coefficients as a function of dimensionless viscosity, $Oh$. 
The ILV scaling prefactor, $C_0$, is of order 1 across the entire range of viscosity (although there is a slight increase in $C_0$ as the viscosity is increased). 
The inertial scaling prefactor, $D_0$, is in good agreement with the value from numerical simulations \cite{Eggers2003}, $D_0=1.62$. 

Figure \ref{rcvsOh}(c) shows the dimensionless crossover time, $\tau_c/\tau_{\text{v}}$, as a function of $Oh$ (where $\tau_{\text{v}}= \mu A/\gamma$ is the viscous timescale). 
Clearly, the accepted formula for the crossover time, $\tau_c/\tau_{\text{v}} \approx Oh^2$, does not agree with the data. 
The measurements are better described by a linear dependence on $Oh$. 

The discrepancy between theory and experiment is also evident in the dimensionless crossover radius, $r_c/A$, versus $Oh$, shown in Fig.\ \ref{rcvsOh}(d). 
The predicted crossover radius is $r_c/A \approx Oh^2$, whereas the data follow $r_c/A \approx Oh$. 
This suggests that the conventional Reynolds number for coalescence, $Re = \rho \gamma r_{\text{min}}/\mu^2$, is wrong.

\subsection{Reynolds number for coalescence}

The viscous-to-inertial crossover can be estimated by the condition that the dimensionless Reynolds number for the flows, $Re=\rho U L/\mu$, is of order unity (where $U$ and $L$ are characteristic velocity- and length-scales in the flows, respectively). 
As was argued in ref.\ \cite{Paulsen2011}, the dominant flows in the viscous-to-inertial crossover correspond to a different Reynolds number than the one used in the literature \cite{Eggers1999, Eggers2003, Wu2004, Yao2005, Thoroddsen2005, Bonn2005}. 
Instead of the conventionally used length scale given by the neck radius, $L=r_{\text{min}}$, a much smaller length scale---the neck height, $r_{\text{min}}^2/A$---describes the size of the flow gradients. 

Paulsen \textit{et al.}\ \cite{Paulsen2011} gave an estimate for the Reynolds number coming from the inertial side of the crossover. 
They found that the crossover time, $\tau_c$, is given by: 
\begin{equation}
\frac{\tau_c}{\tau_{\text{v}}} \approx \frac{64}{D_0^6} \left(\frac{\mu}{\sqrt{\rho \gamma A}}\right) = \frac{64}{D_0^6} \mbox{ } Oh,
\label{ourtc}
\end{equation}
\noindent which is written here using the viscous timescale, $\tau_{\text{v}}$, and the Ohnesorge number. 
Figure \ref{rcvsOh}(c) shows that this prediction is consistent with the crossover times measured in this work.

A similar argument can be made coming from the viscous side of the crossover, which is presented here. 
In the early (ILV) regime, the characteristic speed of the flows is $U=\gamma/\mu$, and the characteristic length-scale is $L=r_{\text{min}}^2/(2A)$, since liquid from each drop moves in to advance the neck. 
Using these scales, the Reynolds number is: $Re=\rho\gamma r_{\text{min}}^2/(2 A \mu^2)$. 
\noindent The dimensionless crossover radius, $r_c/A$, is obtained by setting $Re=1$: 
\begin{equation}
\frac{r_c}{A}\approx \sqrt{2}\left(\frac{\mu}{\sqrt{\rho \gamma A}}\right)=\sqrt{2}\mbox{ }Oh.
\label{rcformula}
\end{equation}

\noindent Figure \ref{rcvsOh}(d) shows that this prediction gives excellent agreement with the data.

In Appendix \ref{prevCross}, I compare the calculated crossover time, Eq.\ (\ref{ourtc}), with previous measurements of the viscous-to-inertial crossover in the literature.

\subsection{High-speed imaging collapse}

High-speed videos of coalescence show that these results also capture the dependance of the crossover on surface tension. 
I coalesce glycerol-water-NaCl mixtures with viscosities ranging from 1.9 to 230 mPa s, and silicone oils with viscosities ranging from 0.82 to 97 mPa s. 
(The silicone oils are electrically insulating, and therefore cannot be measured with the electrical method.) 
Using these liquids, the surface tension is varied by a factor of 5. 
The liquids have $Oh<1$ so that the behavior should be described by the ILV regime and the inertial regime, but not the Stokes regime. 

\begin{figure}[bt] 
\centering 
\begin{center} 
\includegraphics[width=3.1in]{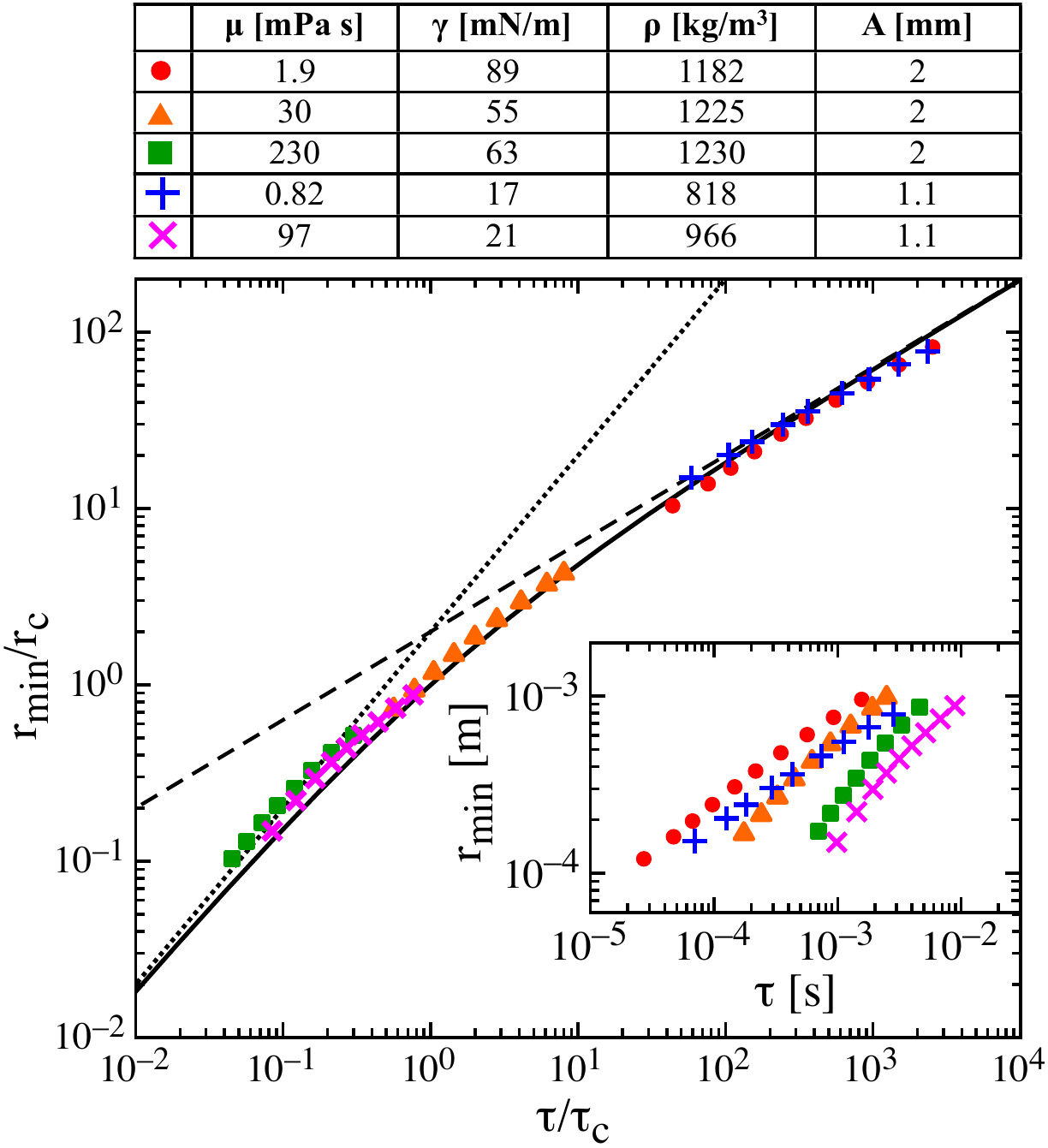} 
\end{center}
\caption{
(Color online)
High-speed imaging of coalescence. 
\textit{Inset:} Neck radius versus time for glycerol-water-NaCl mixtures ($\mu=1.9$, 30, and 230 mPa s) and silicone oils ($\mu=0.82$ and 97 mPa s). 
Other parameters are listed in the legend. 
\textit{Main:} The data is collapsed by rescaling the axes with the crossover radius, $r_c$ (calculated with Eq.\ (\ref{rcformula})), and the crossover time, $\tau_c$ (calculated with Eq.\ (\ref{ourtc})). 
The rescaled data are consistent with Eq.\ (\ref{interpolate}) (solid line). 
} 
\label{videoCollapse}
\end{figure}

The inset of Fig.\ \ref{videoCollapse} shows $r_{\text{min}}$ versus $\tau$ for these liquids. 
When the axes are rescaled with $r_c$ (given by Eq.\ (\ref{rcformula})) and $\tau_c$ (given by Eq.\ (\ref{ourtc})), the data collapse to a master curve, shown in Fig.\ \ref{videoCollapse}. 
The collapsed data follow Eq.\ (\ref{interpolate}), and therefore fall on the electrical data collapse, Fig.\ \ref{muCollapse}. 
These experiments further solidify the new phase diagram for coalescence, shown in Fig.\ \ref{phaseDiagram}(b).

\section{Dynamics of drops during approach}
\label{approach}

\subsection{Drop deformation}

The experiments in this work were performed at ambient air pressure. 
Because the drops approach at finite speed, they can be deformed by the viscous stresses in the air layer between them \cite{Neitzel2002}. 
This deformation could affect the subsequent coalescence dynamics. 
Previous experiments \cite{Case2008,Case2009} using the same electrical method suggest that deformation may be present for approach speeds as low as $10^{-4}$ m/s. 

Here, aqueous NaCl drops are coalesced in air at an approach speed that is varied over $7$ orders of magnitude down to $17$ nm/s, to examine the effects of the ambient gas during approach. 
To achieve constant approach-speeds lower than $10^{-3}$ m/s, a variable-speed syringe pump was used with a wide range of syringe sizes. 
The approach speed, $U_{\text{app}}$, was calculated based on the geometry and the flow rate. 
The coalescence cell was fixed to a vibration-isolation table to suppress disturbances on the drops. 
For high approach-speeds, a gravity-fed system was used: the bottom drop was fed by a reservoir held at a variable height above the coalescence cell, so that hydrostatic pressure caused the bottom drop to grow and impact the top drop. 
For the gravity-fed system, $U_{\text{app}}$ was measured directly with a high-speed camera. 

\begin{figure}[bt] 
\centering 
\begin{center} 
\includegraphics[width=3.2in]{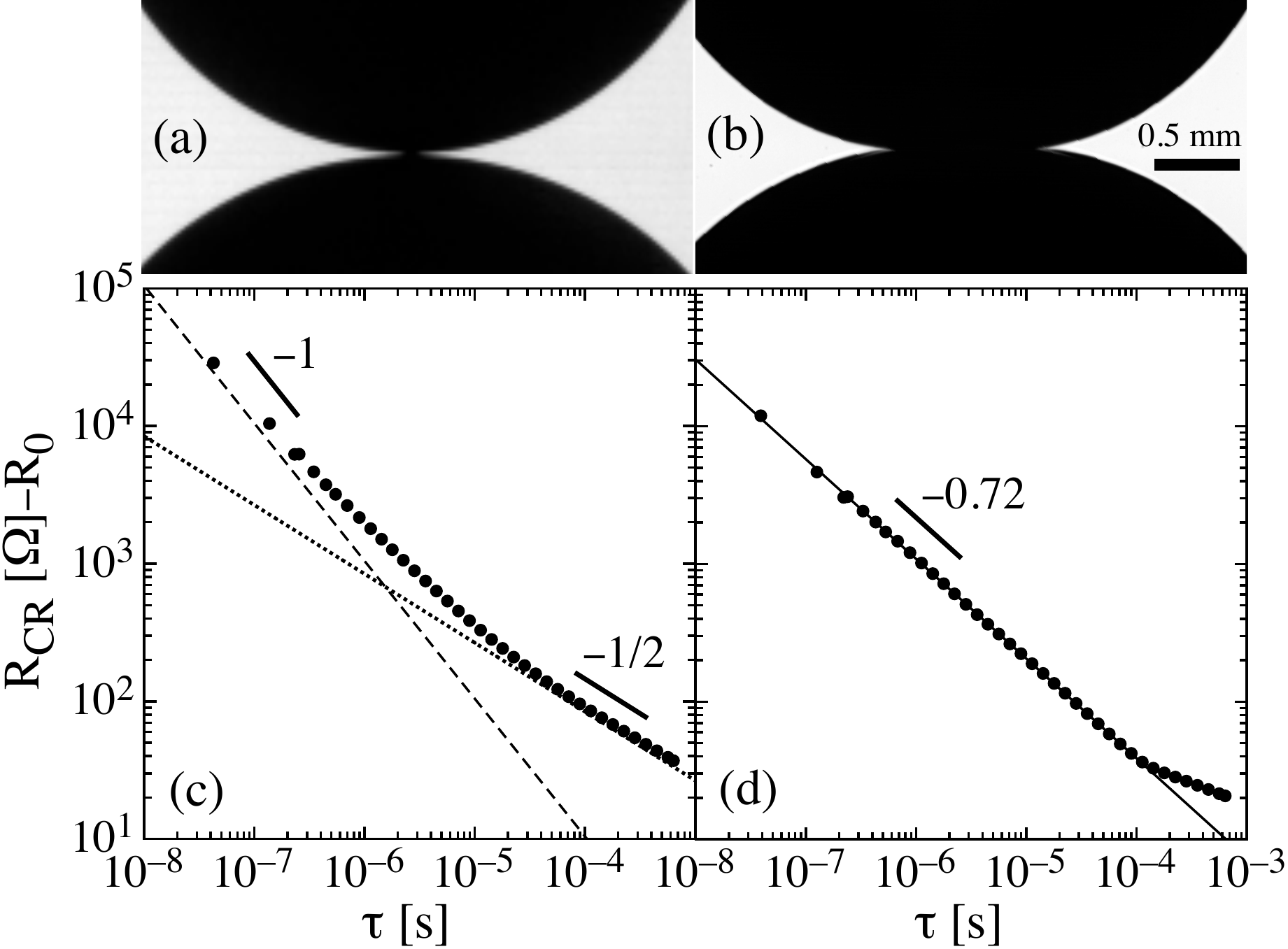} 
\end{center}
\caption{
Optical and electrical measurements of coalescence at low and high approach-speed. 
(a,b) Visual indications of drop deformation. 
Drops are shown within one frame of $t_0$ at low approach-speed: (a) $U_{\text{app}}=8.8\times 10^{-5}$ m/s, and high approach-speed: (b) $U_{\text{app}}=3.3\times 10^{-2}$ m/s. 
At low approach-speed, the drops appear to coalesce as undeformed spheres, whereas the high approach-speed drops are flattened. 
(c,d) Electrical measurements corresponding to the experiments shown in (a,b). 
$R_{\text{CR}}-R_0$ versus $\tau$, where $R_0=1/(\sigma \pi A)$. 
At low approach-speed (c), the resistance follows $\tau^{-1}$ (ILV scaling, dashed line) at early times and $\tau^{-1/2}$ (inertial scaling, dotted line) at late times. 
At high approach-speed (d), the resistance follows $\tau^{-0.72}$ at early times (solid line). 
}
\label{RCRva}
\end{figure}

Figure \ref{RCRva}(a) shows an image taken within one frame of $t_0$ for $U_{\text{app}}=8.8\times 10^{-5}$ m/s. 
The drops appear to be undeformed at the moment of contact. 
At much higher approach-speed, the drops visibly deform before they merge, as shown in Fig.\ \ref{RCRva}(b), for $U_{\text{app}}=3.3\times 10^{-2}$ m/s. 
This transient non-coalescence is due to the pressure provided by the lubricating air layer between the drops. 
Although the drops appear undeformed in the low approach-speed case shown in Fig.\ \ref{RCRva}(a), the image does not rule out the possibility of a small flattened region at the drop tips. 

The electrical method was used to access these small scales. 
Figure \ref{RCRva}(c) shows electrical measurements of the coalescing drops for the low approach-speed case. 
The data follow the behavior shown in earlier sections of this work (e.g., Fig.\ \ref{Schematic}(d)). 
However, for the high approach-speed case, the electrical measurements are qualitatively different, as shown in Fig.\ \ref{RCRva}(d). 
At early times, the resistance appears to follow an approximate power-law with a scaling exponent of $-0.72$. 
At late times, there is an abrupt crossover out of this scaling. 

\begin{figure}[bt] 
\centering 
\begin{center} 
\includegraphics[width=2.7in]{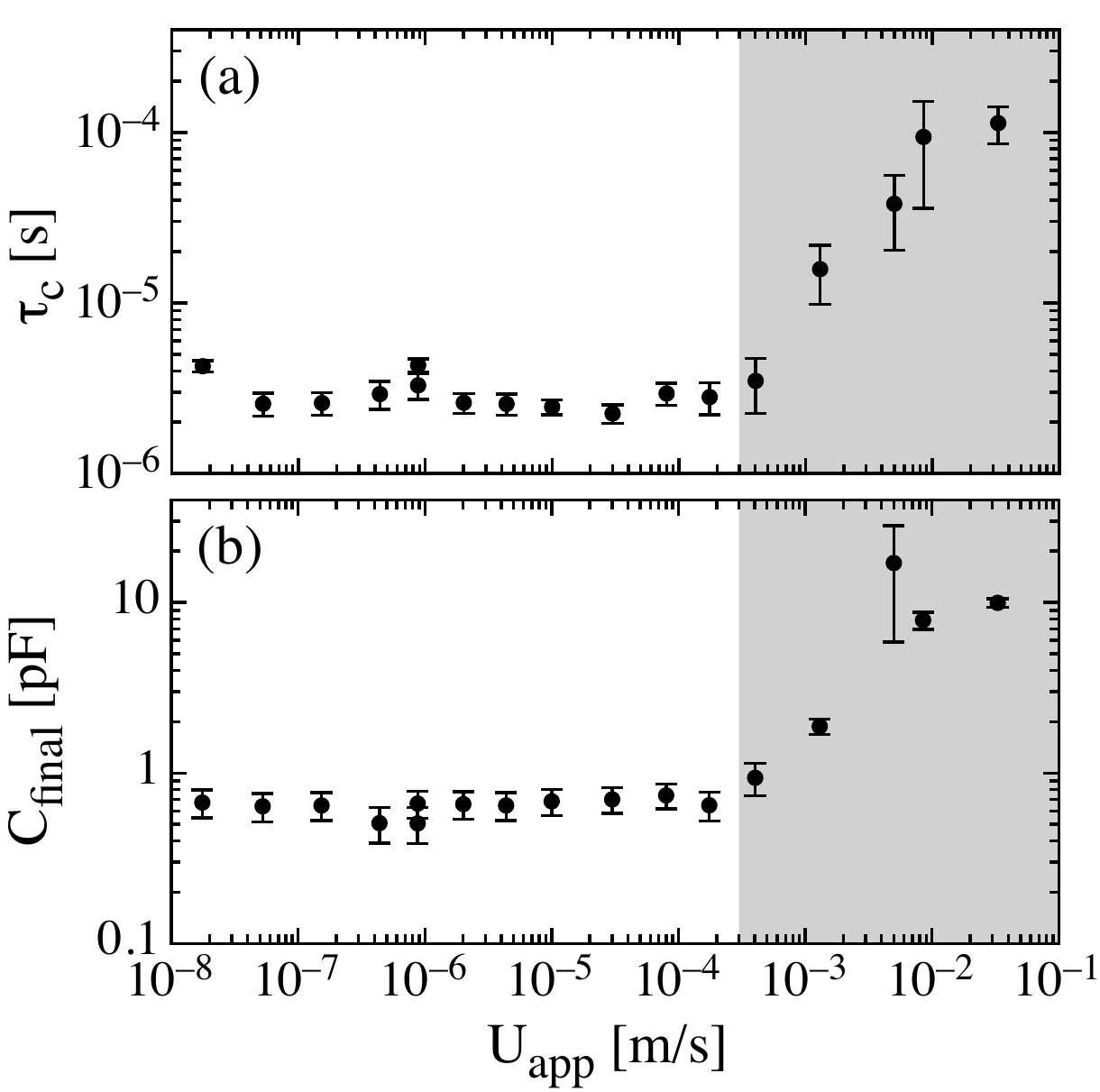} 
\end{center}
\caption{
(a) Crossover time between the early and late electrical-resistance scalings versus approach speed. 
The crossover time is constant for $U_{\text{app}}<U_{\text{app}}^*$, where $U_{\text{app}}^*=(3 \pm 1)\times 10^{-4}$ m/s. 
When $U_{\text{app}}>U_{\text{app}}^*$ (shaded region), $\tau_c$ depends on the approach speed, and is delayed as $U_{\text{app}}$ increases. 
(b) $C_{\text{final}}$ versus approach speed. 
$C_{\text{final}}$ is constant for $U_{\text{app}}<U_{\text{app}}^*$. 
For $U_{\text{app}}>U_{\text{app}}^*$ (shaded region), $C_{\text{final}}$ increases with $U_{\text{app}}$, consistent with an increase in the flattening of the drops before they touch. 
The data are averaged over 800 samples within the final 100 $\mu$s before $t_0$, and $V_{\text{in}} \leq 275$ mV.  
} 
\label{Uapp}
\end{figure}

A crossover time, $\tau_c$, is measured by fitting the early- and late-time data to separate power-laws and determining the point of intersection of the fits. 
(This criteria is equivalent to fitting to Eq.\ (\ref{interpolate}) if the two scalings are the ILV and inertial scalings.) 
Figure \ref{Uapp}(a) shows $\tau_c$ versus approach speed. 
The crossover time is insensitive to the drop approach-speed for $U_{\text{app}}<3 \times 10^{-4}$ m/s. 
For $U_{\text{app}}>3 \times 10^{-4}$ m/s, the crossover time increases approximately linearly with $U_{\text{app}}$, which is correlated with an increase in flattening in the high-speed videos. 
A threshold approach-speed, $U_{\text{app}}^*=(3 \pm 1)\times 10^{-4}$ m/s, separates the two behaviors. 

The capacitance of the drops at the moment of contact, $C_{\text{final}}\equiv C_{\text{CR}}(\tau=0)$, should be sensitive to the amount of drop deformation as well. 
In particular, $C_{\text{final}}$ should grow with the area of the deformed region. 
Fig.\ \ref{Uapp}(b) shows $C_{\text{final}}$ versus approach speed. 
The capacitance shows two behaviors, which fall on either side of the threshold approach-speed, $U_{\text{app}}^*$. 
At high approach-speed ($U_{\text{app}}>U_{\text{app}}^*$), $C_{\text{final}}$ increases with $U_{\text{app}}$ as the drops are increasingly deformed. 
At low approach-speed ($U_{\text{app}}<U_{\text{app}}^*$), $C_{\text{final}}$ is independent of $U_{\text{app}}$, which is consistent with a picture where the drops are undeformed. 

The crossover time and capacitance measurements suggest that drop deformation is absent at low approach-speed, since in the flattening scenario, one expects the amount of flattening to change with $U_{\text{app}}$ \cite{Law2007}. 
In general, $U_{\text{app}}^*$ may depend on the drop size, surface tension, the ambient fluid viscosity, and the drop viscosity. 

\subsection{Inception of coalescence}

\begin{figure}[bt] 
\centering 
\begin{center} 
\includegraphics[width=3.2in]{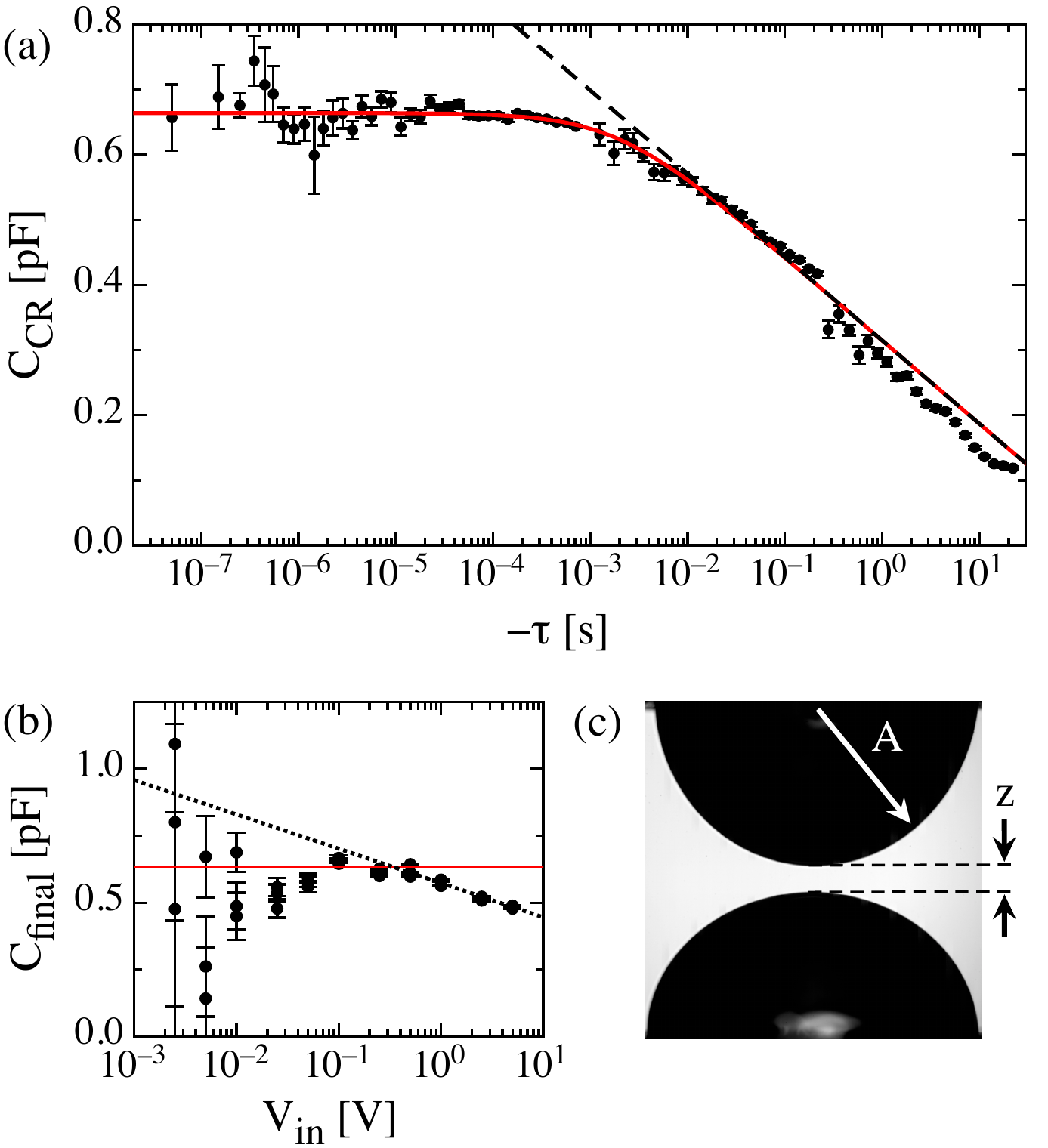} 
\end{center}
\caption{
(Color online)
Drop capacitance before coalescence for drops approaching at speed $U_{\text{app}}=8.8\times 10^{-5}$ m/s $<U_{\text{app}}^*$. 
(a) Capacitance of the coalescence region, $C_{\text{CR}}$, versus time remaining until coalescence, $-\tau$, with $V_{\text{in}}=275$ mV and $f=10$ MHz. 
The drops initiate contact at finite separation, shown by the excellent fit to Eq.\ (\ref{Ceq}) with $z=-U_{\text{app}} \tau+z_0$ using $z_0=160$ nm (solid line). 
For comparison, the same equation is plotted with $z_0=0$ (dashed line). 
The data is averaged and logarithmically binned over 12 experiments. 
(b) $C_{\text{final}}$ versus applied voltage, $V_{\text{in}}$. 
At high voltage, the data is consistent with coalescence initiating when the peak electric field between the drops reaches a threshold value (dotted line: Eq.\ (\ref{Ceq}) with $z_0=V_{\text{in}}/E_{\text{thresh}}$ using $E_{\text{thresh}}=1.2$ MV/m). 
At lower voltage, $C_{\text{final}}$ is consistent with a constant value that is independent of $V_{\text{in}}$ (solid line: $C_{\text{final}}=0.63 \pm 0.05$ pF). 
Each data point is averaged over 800 samples within the final 100 $\mu$s before $t_0$. 
Three measurements were made at each voltage to show the run-to-run variation. 
(c) Approach geometry: two drops of radius $A$ separated by distance $z$. 
} 
\label{CfinalVa}
\end{figure}

To probe the dynamics up to the moment of contact, I measure the capacitance of the coalescence region, $C_{\text{CR}}$, during approach at low speed ($U_{\text{app}}<U_{\text{app}}^*$). 
The separation of the drop surfaces is denoted by $z$ (see Fig.\ \ref{CfinalVa}(c)). 
The mutual capacitance of the two hemispherical drops at small separation ($z\ll A$) should be comparable to the capacitance of two conducting spheres, since the surfaces in close proximity contribute the most charge. 
The latter arrangement can be solved exactly as a series expansion. 
For small $z/A$, the capacitance is given to a good approximation \cite{Love1979} by: 
\begin{equation}
C_{\text{spheres}}(z)\approx \pi\epsilon_0 A \left[\ln\left(\frac{A}{z}\right)+2.54\right].
\label{Ceq}
\end{equation}

Figure \ref{CfinalVa}(a) shows the capacitance of the coalescence region, $C_{\text{CR}}$, measured as a function of the time remaining until coalescence, $-\tau$, for aqueous NaCl drops approaching at low speed. 
The capacitance grows logarithmically with $-\tau$ at first but does not diverge at $\tau=0$. 
This behavior is consistent with the drops initiating contact at a finite separation, $z_0$. 
Plugging $z=-U_{\text{app}} \tau+z_0$ into Eq.\ (\ref{Ceq}) and using $z_0=160$ nm gives excellent agreement with the data. 


Despite the low applied voltage, the electric field between the drops can be very strong when the drop separation is small. 
This is apparent when estimating the peak magnitude of the electric field between the drops during one AC cycle as $E_{\text{max}}\approx V_{\text{in}}/z$. 
One expects the large electric field to attract the drops towards each other when they are very close, thus promoting coalescence and distorting the drop shapes before contact. 

To test the effect of the applied voltage on the initiation of coalescence, $C_{\text{final}}$ was measured as a function of applied voltage, $V_{\text{in}}$. 
Figure \ref{CfinalVa}(b) shows the results for $V_{\text{in}}$ ranging from 2.5 mV to 5 V. 
For $V_{\text{in}}>0.3$ V, the data are consistent with the drops forming a connecting neck when the intervening electric field exceeds a threshold value, $E_{\text{thresh}}$. 
The data are fit well by $E_{\text{thresh}}=1.2 \pm 0.2$ MV/m, using Eq.\ (\ref{Ceq}) for the capacitance of the drops as a function of the final separation, $z_0$, and substituting $z_0\approx V_{\text{in}}/E_{\text{thresh}}$. 
While this value is only slightly smaller than the approximate dielectric strength of air at large distances (3 MV/m), the dielectric strength of air at these short distances is much greater \cite{Townsend1915}. 

Having argued that dielectric breakdown does not occur, I now address whether the applied voltage deforms the drops. 
A recent study measured the deformation of two nearby drops with an applied DC electric potential difference \cite{Bird2009}. 
Their experiments showed that the drops sharpen into cones, and they measured a cone angle of roughly $20^{\circ}$ for a potential difference of $500$ V (where $0^{\circ}$ corresponds to no deformation). 
Their measurements of the cone angle are approximately linear for electric potentials between $0$ and $500$ V, suggesting that the cone angle would be less than $0.08^{\circ}$ for the applied voltages used in this work ($V_{\text{in}}\leq 2$ V). 
(The angle is likely diminished even further since the measurements in this work are AC instead of DC.) 
Thus, any deformation of the drops is expected to be on a small scale, although it could contribute to forming the initial microscopic neck for $V_{\text{in}}>0.3$ V. 

Figure \ref{CfinalVa}(b) shows that at lower voltages, $V_{\text{in}}<0.3$ V, $C_{\text{final}}$ is roughly constant within error, and the description invoking a threshold electric field is a poor fit. 
Instead, the data are consistent with a picture where Van der Waals forces initiate coalescence at finite separation when $V_{\text{in}}$ is small. 
For the data at low voltages, I measure $C_{\text{final}}=0.63 \pm 0.05$ pF, giving a best fit of $z_0=280$ nm. 
Because the capacitance is logarithmic in drop separation, the experimental error on $z_0$ is large; the data are consistent with $z_0$ ranging from $120$ nm to $650$ nm. 


\subsection{Initial neck size}

Finally, I address the finite length and width of the liquid neck that is formed at the inception of coalescence. 
Due to the finite separation of the drops at the moment of contact, the separation between the drop interfaces at radius $r$ will be given by $r^2/A+z_0$, instead of simply $r^2/A$ as was assumed in previous sections. 
However, this correction becomes relatively smaller as $r_{\text{min}}$ grows. 
Numerical simulations \cite{Baroudi2013} where low-viscosity drops initiate contact by forming a small fluid neck at finite separation show that after a short delay, the dynamics converge onto the predicted scaling (i.e. Eq.\ (\ref{invScaling})). 

Previous high-speed imaging studies have reported values for the initial finite radius of the fluid neck (referred to as $r_0$) at the inception of liquid drop coalescence in air. 
Values reported were $r_0=50$ $\mu$m for $U_{\text{app}} \lesssim 0.1$ mm/s (ref.\ \cite{Thoroddsen2005}) and $r_0=43.8 \pm 4.3$ $\mu$m for $U_{\text{app}}=6.6$ mm/s (ref.\ \cite{Fezzaa2008}). 
In contrast, I measure $r_{\text{min}}$ down to $0.7$ $\mu$m at $\tau=50$ ns for aqueous NaCl drops at low approach-speed. 
This is a significantly smaller upper bound for the initial size of the neck for aqueous NaCl drops at low approach-speed. 

\section{Conclusion}

In summary, I have presented supporting evidence for the new phase diagram for liquid drop coalescence in vacuum or air, developed in refs.\ \cite{Paulsen2012, Paulsen2011}. 
The theoretically unanticipated inertially limited viscous regime was found to be the true asymptotic regime of coalescence for drops of any finite viscosity. 
In this regime, surface-tension, viscosity, and inertia all balance. 
Viscous drops ($Oh>1$) transition into the Stokes regime once the neck is sufficiently large to pull the drops towards each other. 
Low-viscosity drops ($Oh<1$) transition into the inertial regime at late times. 

In the inertially limited viscous regime and the Stokes regime, the center-of-mass motion of the drops was found to track with the motion of the backs of the drops, further solidifying the force balance argument that identified the inertially limited viscous regime in ref.\ \cite{Paulsen2012}. 
This work provides similarity solutions for the neck shapes in these two regimes, and the new phase diagram for coalescence was shown to apply for different boundary conditions. 

Additional evidence was provided for the surprisingly late viscous-to-inertial crossover (from the inertially limited viscous regime to the inertial regime), including an alternative method of data-collapse, a Reynolds-number argument coming from the viscous side, and high-speed imaging experiments where the surface tension was varied. 
The agreement of the new coalescence Reynolds number with the data supports the new picture for the flows, which must have a significant gradient on a small axial length scale set by the neck height, $r_{\text{min}}^2/A$. 

Many of the results are based on electrical measurements, which were shown to have an insignificant effect on the coalescence dynamics reported here. 
At low approach-speed and low applied-voltage, the drops coalesce at finite separation as undeformed spheres. 

Whereas this work has established the behavior of liquid drop coalescence in vacuum or air, further work is needed to determine how an outer fluid with significant density or viscosity alters the coalescence phase diagram. 


\begin{acknowledgments} 
I thank Sidney Nagel and Justin Burton for their guidance, support, and keen insight throughout this work. 
I am also particularly grateful to Santosh Appathurai, Osman Basaran, Sarah Case, Thomas Rosenbaum, Savdeep Sethi, and Wendy Zhang. 
I thank Michelle Driscoll, Efi Efrati, Nathan Keim, and Tom Witten for their assistance and for many illuminating discussions. 
This work was supported by NSF Grant DMR-1105145 and by NSF MRSEC DMR-0820054. 
\end{acknowledgments}

\appendix

\section{Checks on the electrical method}
\label{elecSystematic}

Here I report checks on the electrical method to test its accuracy, and to assess whether the applied voltage alters the observed coalescence dynamics. 
In addition, several checks on the electrical method were performed by Case \textit{et al.}\ \cite{Case2009}, including varying the applied voltage, varying the DC component of the applied voltage, and varying the ionic concentration of the drops. 

As a basic check of the accuracy of the electrical measurements, the coalescence cell was replaced with known circuit elements spanning the range of impedances observed in coalescence. 
Their impedances were found to be within error bars of their nominal values. 
As a dynamical check on the electrical method, I measured a pinch-off event of aqueous NaCl. 
Following the analysis in refs.\ \cite{Case2009,Burton2004}, I found the same power-law scaling for the neck radius as had been found by other methods \cite{Basaran2002}. 

\begin{figure}[bt] 
\centering 
\begin{center} 
\includegraphics[width=3.3in]{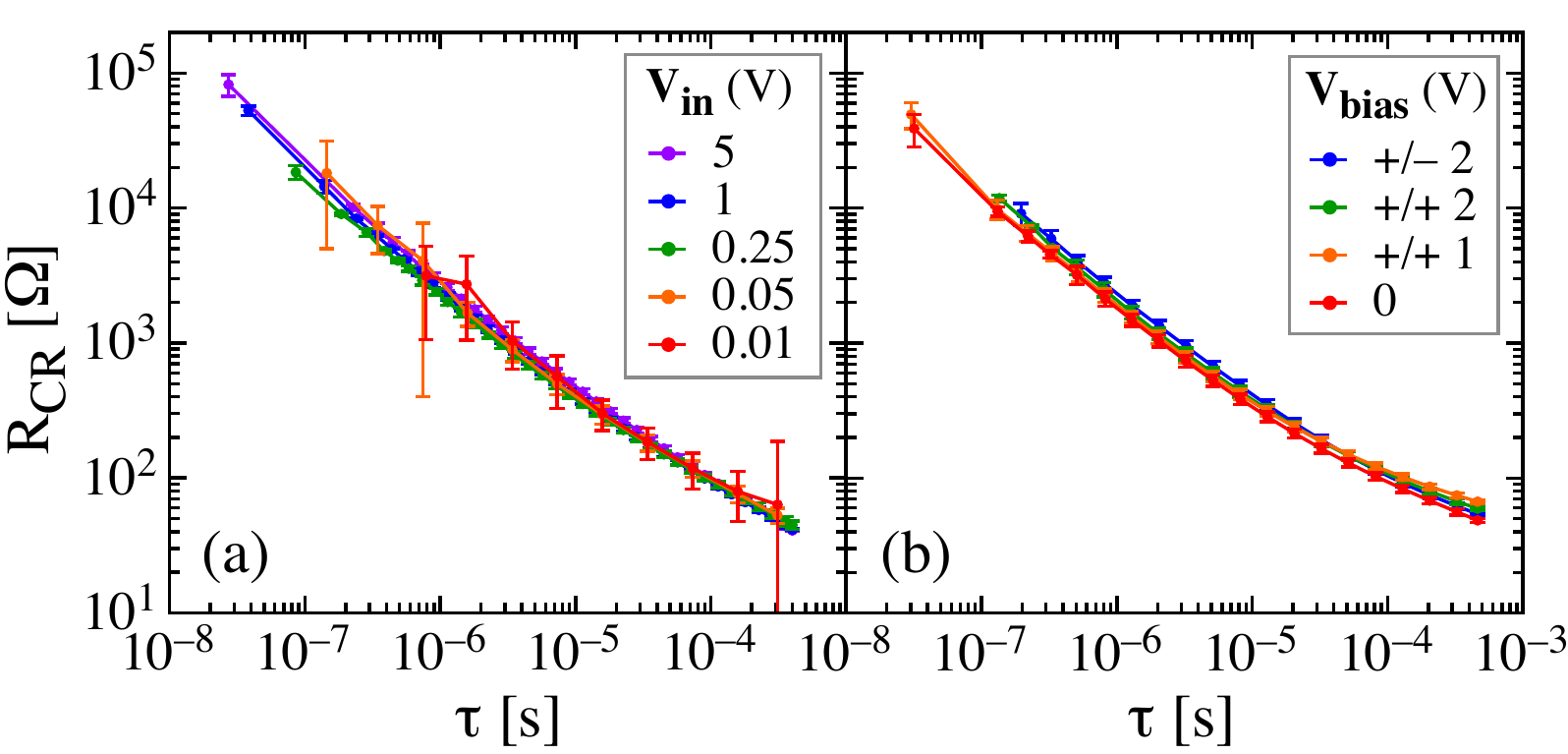} 
\end{center}
\caption{
(Color online)
Checks on electrical measurements of $R_{\text{CR}}$ versus $\tau$. 
The data for each set of parameters are binned and averaged over 3 coalescences. 
(a) The driving voltage, $V_{\text{in}}$, is varied from $10$ mV to $5$ V. 
There is no significant change in the measurements. 
(b) A DC bias voltage is applied to each drop during the electrical measurement, with magnitude $V_{\text{bias}}$ and with opposite ($+/-$) or matching ($+/+$) polarity on the two drops. 
Here, $V_{\text{in}}=275$ mV. 
The results are within error bars of each other. 
}  
\label{elecCheck}
\end{figure}

To address whether the applied voltage alters the coalescence dynamics, I varied the voltage amplitude, $V_{\text{in}}$, over a large range, and compared measurements of the resistance of the coalescence region, $R_{\text{CR}}$, versus time. 
This check had been performed in ref.\ \cite{Case2009}, where $V_{\text{in}}$ was varied from 25 mV to 1 V. 
Here, a broader range of $V_{\text{in}}$ was tested, from 10 mV to 5 V. 
As shown in Fig.\ \ref{elecCheck}(a), measurements of $R_{\text{CR}}$ are within error bars of each other over this range. 

Even when $V_{\text{in}}$ is small, the electric field in the region between the drops becomes large as the drops approach contact. 
It has been suggested that an applied voltage is not a proper method for detecting the rupture of the ambient fluid between approaching drops, since the drops are deformed due to the high electric field between them \cite{Lukyanets2008}. 
To address this issue, I performed tests where I altered the electric fields between the drops by applying a DC bias voltage to each drop. 
Blocking capacitors were added in series on each side of the cell to pass through the AC measurement signal. 
Configurations were tested where the two drops were given a bias of equal or opposite polarity, as well as cases where $V_{\text{bias}}>V_{\text{in}}$. 
The results are shown in Fig.\ \ref{elecCheck}(b). 
Measurements of $R_{\text{CR}}$ are within error bars of each other over all of the experiments. 

I make note of one experimental peculiarity, observed previously in ref.\ \cite{Case2009}, that remains unexplained. 
In the experiments, a small DC spike ($\lesssim 30$ mV across the digitizer input) occurred at $t_0$, with opposite polarity in either channel of the digitizer. 
This spike was still present when the coalescence cell was not connected to the function generator. 
Although it is still unexplained, the spike does not affect the results, since any DC component of the output signal is removed in the analysis.

\section{Previous viscous-to-inertial crossover measurements}
\label{prevCross}

Three other studies have reported measurements of viscous-to-inertial crossover in liquid drop coalescence \cite{Thoroddsen2005, Bonn2005, Burton2007}. 
Here, I briefly discuss how their crossover-time measurements compare to the present work. 

First, in ref.\ \cite{Bonn2005}, the authors reported a crossover time of $0.45\pm 0.15$ ms for $50$ mPa s silicone oil, which they remark is in agreement with the prediction of $0.32$ ms given by the conventional coalescence Reynolds number. 
For the fluid parameters in that study, the present work predicts $\tau_c\approx (64/D_0^6) \mu^2\sqrt{A/\rho \gamma^3}=4.5$ ms. 
Although the conventional Reynolds number agrees with their result, the crossover was identified as the point where the neck radius versus time first departed from a linear scaling. 
This method identifies a time that is roughly an order of magnitude smaller than $\tau_c$, due to the large width of the crossover region (see Fig.\ \ref{muCollapse}). 


Second, a high-speed imaging study \cite{Thoroddsen2005} reported a crossover time of $2.5$ ms in a viscous glycerol-water mixture, with $\mu=493$ mPa s. 
For their fluid parameters and drop sizes, the Ohnesorge number is equal to $1.4$, so that a transition into the Stokes regime (instead of the inertial regime) should occur in the last moments of merging. 
However, the departure from the linear scaling is likely due to finite-size effects; the neck expansion slows when the fluid neck becomes close to the size of the drops. 

Finally, a viscous-to-inertial crossover was reported in ref.\ \cite{Burton2007} for the coalescence of quasi-2D liquid alkane lenses floating on water. 
They saw a crossover length of $r_c=250$ $\mu$m, in stark contrast to the conventionally assumed crossover length, $r_c=\mu^2/\rho \gamma$, which is equal to $0.5$ $\mu$m for their system (where $\gamma$ is the line tension). 
They suggested that the crossover should occur when the velocities from the viscous and inertial scalings are equal, which gives $r_c=2\pi\mu\sqrt{A/\rho \gamma}=253$ $\mu$m. 
This was close to the experimental observation. 
The new coalescence Reynolds number presented in ref.\ \cite{Paulsen2011} and argued for in this paper puts their argument on a more solid fluid-dynamical footing. 
Using Eq.\ (\ref{rcformula}), the expected crossover length for their system is $r_c=\sqrt{2} Oh A=\mu \sqrt{2A/ \rho \gamma}=$ 59 $\mu$m. 
Although this calculation is smaller than what they measure, their quasi-2D system has additional dissipation in the water subphase, which should increase the prefactor for the viscous term in the Reynolds number and therefore delay the crossover.

\bibliography{CoalescenceBib}

\end{document}